%% file: main.tex
\documentclass[journal]{IEEEtran}
\IEEEoverridecommandlockouts
\usepackage[noadjust]{cite}
\usepackage{amsmath,amssymb,amsfonts}
\usepackage{algorithmic}
\usepackage{graphicx}
\usepackage{xcolor}

\usepackage{moreverb}
\usepackage{url}
\usepackage[
    colorlinks,
    bookmarksopen,
    bookmarksnumbered,
    citecolor=black,
    urlcolor=black,
    allcolors=black
]{hyperref}

\usepackage{amssymb}
\usepackage{booktabs}
\usepackage[inline]{enumitem}
\usepackage{fontawesome5}
\usepackage[numbers]{natbib}
\usepackage{multirow}
\usepackage{pgfplots}
\usepackage{pgfplotstable}
\usepackage{sansmath}
\usepackage{subcaption}
\usepackage{siunitx}
\usepackage{svg}
\usepackage{tabularx}
\usepackage{tikz}
\usepackage{xspace}
\usepackage{ragged2e}

\input{colors}

\newcommand\fancyname{BEAST\xspace}

\def\largejup{JUPITER-2.4B-64K}
\def\smalljup{JUPITER-0.7B-1.8K}
\def\largelumi{LUMI-1.1B-64K}

\pgfplotsset{
    compat=newest,
    gbplot/.style={
        axis x line=bottom,
        axis y line=left,
        enlarge x limits=0.05,
        label style={
            font=\footnotesize\bfseries
        },
        ticklabel style={
            font=\sansmath\scriptsize\sffamily,    
            /pgf/number format/.cd,
            fixed,
            precision=2,
        },
        scaled ticks=false,
        grid=major,
        grid style={ 
            draw=hgfgray,
            densely dotted
        },
        legend style={
            draw=none,
            inner sep=0,
            font=\footnotesize\sffamily,
            anchor=north, 
            row sep=0.2em,
            /tikz/every even column/.append style={
                column sep=1.0em
            }
        },
        legend cell align=left,
        axis on top
    }
}
\usepgfplotslibrary{fillbetween}
\tikzset{every picture/.style={/utils/exec={\sffamily}}}

\DeclareSIUnit\gpu{GPU}
\DeclareSIUnit\node{Node}
\DeclareSIUnit\flop{FLOP}
\DeclareSIUnit\flops{\flop\per\second}
\DeclareSIUnit\flopsnode{\flops\per\node}
\DeclareSIUnit\sample{sample}
\DeclareSIUnit\samplepsec{\sample\per\second}
\DeclareSIUnit\forecast{forecast}
\DeclareSIUnit\Forecast{Forecast}
\DeclareSIUnit\forecastpsec{\forecast\per\second}
\DeclareSIUnit\Forecastpsec{\Forecast\per\second}
\usetikzlibrary{
    3d,
    arrows.meta,
    backgrounds,
    calc,
    ext.paths.ortho,
    patterns,
    perspective,
    positioning
}

\def\BibTeX{{\rm B\kern-.05em{\sc i\kern-.025em b}\kern-.08em
    T\kern-.1667em\lower.7ex\hbox{E}\kern-.125emX}}
\begin{document}

\title{4D Parallelism Unlocks Exascale Bayesian Neural Networks for High-Fidelity Atmospheric Modeling\\

}
\DeclareRobustCommand*{\IEEEauthorrefmark}[1]{%
  \raisebox{0pt}[0pt][0pt]{\textsuperscript{\footnotesize\ensuremath{#1}}}}

\author{\IEEEauthorblockN{
Deifilia Kieckhefen\IEEEauthorrefmark{1}$^\dagger$, 
Juan Pedro Gutiérrez Hermosillo Muriedas\IEEEauthorrefmark{1}$^\dagger$,
Lars Helge Heyen\IEEEauthorrefmark{1},
Mathis Bode\IEEEauthorrefmark{2},
Iida Hakulinen\IEEEauthorrefmark{3},
Andreas Herten\IEEEauthorrefmark{2},
Chelsea Maria John\IEEEauthorrefmark{2},
Thorsten Kurth\IEEEauthorrefmark{4},
Anni Moisala\IEEEauthorrefmark{3},
Asena Karolin Özdemir\IEEEauthorrefmark{1},
Kaleb Phipps\IEEEauthorrefmark{1},
Oskar Taubert\IEEEauthorrefmark{3},
Arvid Weyrauch\IEEEauthorrefmark{1},
Markus Götz\IEEEauthorrefmark{1}, and
Charlotte Debus\IEEEauthorrefmark{1}
}\\[0.5em]
\IEEEauthorblockA{\IEEEauthorrefmark{1}Karlsruhe Institute for Technology, Germany}\\
\IEEEauthorblockA{\IEEEauthorrefmark{2}Forschungszentrum Jülich, Germany}\\
\IEEEauthorblockA{\IEEEauthorrefmark{3}CSC – IT Center for Science Ltd., Finnland}\\
\IEEEauthorblockA{\IEEEauthorrefmark{4}NVIDIA, Switzerland}
\thanks{\noindent$^\dagger$These authors contributed equally.\\ Corresponding author: C. Debus (charlotte.debus@kit.edu)}
}

\maketitle

\begin{abstract}
We present \fancyname, the first-ever Bayesian Swin Transformer for atmospheric forecasting on \ang{0.25} global resolution able to accurately quantify both aleatoric and epistemic uncertainty. 
To overcome the associated computational bottlenecks, we devise an orthogonal 4D-parallelization scheme that introduces a unique domain-tensor-parallelism strategy and a novel uncertainty parallel method, enabling us to fully leverage GPU capacity and efficiently scale model training.
For a \num{2.4}-billion-parameter model, we achieve a peak performance of \qty{3.96}{\exa\flops} on \num{20480} NVIDIA GH200 GPUs on the JUPITER supercomputer. 
We train \fancyname~as a \num{700}-million-parameter model with \num{96} random weight samples on \num{384} nodes on \num{40} years of data for nearly one million gradient updates. This model achieves predictive skill scores competitive with state-of-the-art probabilistic atmospheric AI models and numerical models, and can predict extreme events with exceptional skill, while generating large ensembles \num{3} to \num{4} times faster than the current-best AI model. 
Our contribution unlocks the potential of high-fidelity uncertainty quantification in atmospheric AI models, heralding a new era for AI-based models in climate and Earth system sciences.

\end{abstract}


\section{Overview}
Artificial intelligence (AI) models for atmospheric forecasting are revolutionizing weather and climate sciences, owing to their promise of generating predictions with skill scores competitive to numerical models at a fraction of the computational resources.
Current atmospheric AI models learn a set of point-value (scalar) weights through iterative optimization over a finite dataset. 
However, this approach disregards the fact that the results of the optimization process can be ambiguous, meaning that the weights could also have different numeric values. From a modeling point of view, this corresponds to the \emph{epistemic uncertainty} of the model -- a systematic error originating from the incompleteness of the training datasets.

The uncertainty in the model weights is dependent on the size and quality of the dataset: the more accurately the training dataset represents the true data distribution of the modeled problem, the more certain the weights are and the better the prediction becomes. For atmospheric AI models, this has two direct consequences. For one, the prediction of rare weather phenomena, such as extreme events, is challenging, as they are sparsely represented in the training data, or not at all. Second, the prediction quality is directly coupled to the size of the dataset~\cite{yu2026scaling}, as more data items typically provide a better approximation of the true data distribution. Hence, modeling rare physical processes of Earth system components for which only small-sized datasets are available remains difficult.

After the recent successes in AI-based weather forecasting with deterministic large-scale neural networks~\cite{lam2023,bi2023accurate,kurth2023fourcastnet,Boris2023SFNO}, probabilistic models for quantifying the uncertainty of a prediction have become the next frontier. Current approaches to uncertainty quantification in AI models focus on the uncertainty of the atmospheric input state to generate ensemble forecasts, similar to ensemble generation via input perturbation in numerical models, as used for HENS~\cite{mahesh2025huge1, mahesh2025huge2}. In particular, diffusion-based methods like GenCast~\cite{price2024probabilistic}, functional generative networks (FGN)~\cite{alet2025skillful}, FourCastNet3 (FCN3)~\cite{bonev2025fourcastnet3geometricapproach}, and AIFS-CRPS~\cite{lang2026aifs} have achieved impressive results. 
However, these existing approaches capture only the so-called \emph{aleatoric} uncertainty, which stems from the inherent chaotic nature of the physical processes in the atmosphere.
In contrast, modeling \emph{epistemic} uncertainty in atmospheric AI models has received little attention so far. FGN and HENS approximate Bayesian inference with deep ensembles~\cite{lee2015m, lakshminarayanan2017simple}, i.e., training multiple models from independently randomly initialized parameters; however, a probabilistic model formulation that captures both epistemic and aleatoric uncertainty has not yet been realized.

Bayesian neural networks (BNNs) provide a principled framework for modeling epistemic uncertainty. BNNs extend classical neural networks by replacing the scalar weights with probability distributions, thereby transforming point predictions into predictive distributions. Compared to alternative methods, like input perturbation or diffusion, BNNs offer a more comprehensive and theoretically grounded approach to uncertainty estimation, leading to better-calibrated and more informative ensemble predictions~\cite{gawlikowski2023survey}.
The distributional parameterization of BNN weights allows for distribution tails to be better captured, leading to more accurate predictions of extreme events.  
Moreover, BNNs can be trained to competitive prediction accuracy on much less data compared to their deterministic (non-Bayesian) counterparts, owing to their capability of accounting for the under-representation of the data distribution in small training datasets~\cite{jospin2022hands}. Hence, they have great potential for AI-based modeling of other components of the Earth system, where training data is much scarcer.

However, training BNNs requires significantly more computational resources than training standard neural networks. 
For one, the probabilistic parameterization of network weights effectively doubles the number of parameters that must be learned, increasing the GPU memory requirements of model training. Second, each forward pass on a data batch has to be performed on a set of random weight samples drawn from the current weight distribution, and the resulting predictions, i.e., the ensemble, have to be aggregated for the backward pass. Evaluation of multiple samples per data item linearly increases computational cost and memory demand. For large-scale atmospheric models with hundreds of millions to billions of parameters, these costs become prohibitive. As a result, large-scale implementations of BNNs in this domain have not yet been realized to date.

We set out to break this barrier and realize the first-ever Bayesian model for atmospheric forecasting at global \ang{0.25} resolution. We strategically accelerate model training at all scales, from optimizing single-node performance by leveraging lower precision data types, just-in-time compilation, and fast GPU-to-GPU communication, to minimizing inter-node communication necessary for Bayesian model training. 
This allows training of a \num{2.4}-billion-parameter Bayesian Shifted Window (Swin) Transformer for \SI{12}{\hour} and \SI{24}{\hour} forecasts at a peak performance of \qty{3.96}{\exa\flops} on \num{5120} nodes on the JUPITER supercomputer.

\section{State of the Art}
Regardless of the application, parallelism and scalability are paramount to enabling the training of large-scale models. The conventional way of accelerating training is \emph{data parallelism} (DP)~\cite{li2020pytorch, ben2019demystifying}, where each process holds a copy of the model weights, independently performs a forward--backward pass of its local subset of the dataset, and computes the corresponding gradients. The gradients are then averaged through an \texttt{allreduce} operation, and the weights are updated locally.
The number of iterations required for an epoch can be reduced by increasing the number of data-parallel processes to increase the global batch size.  While data parallelism scales well in a computational manner, having large global batch sizes can cause models to generalize poorly, ultimately limiting scalability~\cite{keskar2017large}.

Recent work on AI models in weather and climate science has focused extensively on so-called foundation models, i.e., very large models that are pre-trained in a self-supervised fashion on large heterogeneous datasets, and subsequently fine-tuned for downstream tasks~\cite{nguyen2023climax, lessig2023atmorep, bodnar2025foundation}. In consequence, model sizes have been growing dramatically beyond the billion-parameter mark. The affirmation of neural scaling laws for AI weather models~\cite{yu2026scaling}, stating that predictive skill improves with quadratically more data and weights, further fuels this trend. As growing model sizes quickly exceed the memory limitations of a single GPU, the field of \emph{model parallelism} has emerged to distribute model weights and associated computations across multiple GPUs. 

There exist multiple forms of model parallelism, as there are different ways (axes) to parallelize the model architecture.
The simplest form of model parallelism is \emph{pipeline parallelism}, where sequential stages, i.e., blocks of layers, are distributed across GPUs~\cite{huang2019gpipe}. While pipeline parallelism circumvents the GPU-memory bottleneck and entails simple communication schemes with only point-to-point operations, the inherent sequentiality of running through the network layers in the forward-backward pass causes GPU idle time, precluding efficient GPU utilization. Another drawback of pipeline parallelism is the necessity to perform asynchronous gradient updates, which degrades predictive performance~\cite{Harlap2018PipeDream}.

An orthogonal approach to pipeline parallelism is \emph{tensor parallelism}, i.e., distributing the weight matrices of each layer across multiple GPUs. 
Megatron-LM~\cite{shoeybi2020megatron} was among the first works to utilize tensor parallelism, efficiently enabling the training of multi-billion-parameter Transformer models by parallelizing across attention heads and feed-forward weight matrices. 
In Transformer models specifically, the quadratic complexity in sequence length of the attention mechanism provides a particular computational burden and memory bottleneck for long sequence lengths. This is oftentimes addressed with \emph{sequence parallelism}~\cite{jacobs2024system}. 

Next to the memory demand of the large weight matrices themselves, the associated gradients and optimizer states can also lead to a severe memory bottleneck. To deal with this, hybrid data--model-parallel approaches have been introduced, such as Zero Redundancy Optimizer (ZeRO)~\cite{rajbhandari2020zero} or Fully Sharded Data Parallelism (FSDP)~\cite{zhao2023pytorchfsdp}.
ZeRO partitions the parameters, activations, and optimizer states across GPUs, and uses \texttt{allgather} operations to reconstruct each layer upon demand. The approach achieves superlinear scaling due to reduced memory footprints. FSDP shards data batches and model parameters. In a given layer, the weights are gathered to compute the full activations. Afterward, the unnecessary activations are discarded to reduce the memory footprint.

The main memory bottleneck for training large language models (LLMs) originates from large weight matrices; in contrast, Vision-Transformer (ViT)-like architectures with long sequence lengths ($\approx10^6$) face a different challenge: large activation memory footprints.
While this can be mitigated by techniques such as gradient offloading to the CPU~\cite{ren2021zero,rajbhandari2020zero} or gradient recomputation~\cite{chen2016training}, these strategies can come at the cost of computational performance. 
Thus, novel parallelization strategies are required to address the significant bottlenecks arising from the size of the data items.

As one of the first works in the field, FourCastNet v1 \cite{kurth2023fourcastnet} employed tensor parallelism on its Adaptive Fourier Neural Operator (AFNO) architecture by applying feature parallelism, splitting the fast Fourier transforms across processes. It also distributes the multi-layer perceptron (MLP) at the end of an AFNO block, sharding model parameters in alternating row and column dimensions. The approach enables scaling up to \num{3808} NVIDIA A100 GPUs, attaining \qty{140.8}{\peta \flops} in mixed precision, which amounts to \qty{11.9}{\percent} of peak performance~\cite{kurth2023fourcastnet}.

ORBIT~\cite{wang2024orbit} used a ViT~\cite{dosovitskiy2020image}, trained on global \ang{1.4} resolution with up to 91 channels, and scaled beyond the 1-billion-parameter model size. ORBIT introduces Hybrid-STOP, which fully shards the model's weights as well as data batches, thus combining tensor parallelism and FSDP. Additionally, communication is optimized by using layer wrapping, i.e., overlapping computation and prefetching of subsequent model layers. Through these means, they train a model with up to 113 billion parameters across \num{49152} GPUs. ORBIT-2 performs a spatial downscaling task, parallelized through the TILES approach ~\cite{wang2025orbit2}. In TILES, the spatial dimension is divided into overlapping spatial tiles, upon which each GPU calculates local attention. With their biggest model of 10B parameters, they achieve a sustained computational throughput of \qty{368}{\peta\flops} on \num{512} nodes (\num{2048} GH200 GPUs) on the Alps system.

AERIS~\cite{hatanpaa2025aeris} scales training of a Swin-Transformer-based~\cite{liu2021swin} diffusion model with $1\times1$ patch embedding up to \num{10080} nodes and utilizes data, pipeline, and Sequence-Window (SWiPe) parallelism. SWiPe partitions the image across windows, and further parallelizes the sequence within each window with DeepSpeed Ulysses's sequence parallelism~\cite{jacobs2024system}. Their biggest model (80B parameters) reaches \qty{10.21}{\exa \flops} on \num{10080} nodes of the Aurora system, corresponding to roughly \num {10} forecasts per second.

\begin{figure*}[ht!]
    \centering
    \input{figures/graphical-abstract.tikz}
    \caption{Schematic of \fancyname's 4D parallelization hierarchy from left to right. Domain‑Tensor‑Parallelism (DTP) decomposes inputs across spatial and channel dimensions. Uncertainty‑Parallelism (UP) runs independent aleatoric noise injections and epistemic weight samplings. Data‑Parallelism (DP) processes separate, non‑overlapping ERA5 data chunks in parallel.}
    \label{fig:graphical-abstract}
\end{figure*}
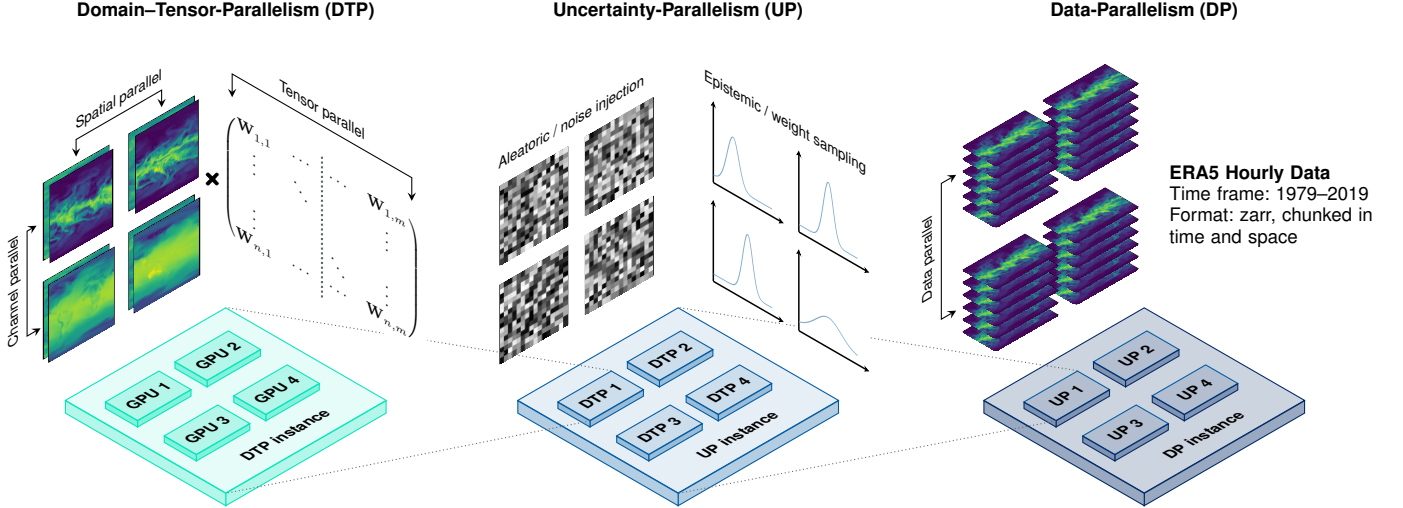
\section{Innovations}
Our main innovations overcome the challenges in training large Bayesian neural networks by introducing a 4-dimensional parallelization strategy to alleviate memory bottlenecks and to maximize per-evaluation throughput. 

The architecture for our \emph{Bayesian Exascale Atmospheric Swin Transformer} (\fancyname) features a 2D-Swin Transformer~\cite{liu2021swin} with one down- and up-sampling layer. 
We train \fancyname according to the Bayesian model formulation
\begin{align*}
    p(y^{*} \mid x^{*}, \mathcal{D})
    = \int p(y^{*} \mid x^{*}, \mathbf{w}) \, 
      p(\mathbf{w} \mid \mathcal{D}) \, d\mathbf{w}
\end{align*}
where $\mathcal{D}$ is the training data, $\mathbf{w}$ are the model weights, $x^*$ is the input and $y^*$ is the output. 
The exact computation of the posterior $p(\mathbf{w} \mid \mathcal{D})$ is generally intractable. To circumvent this, variational inference~\cite{graves2011advances, hoffman2013stochastic} approximates the posterior by a known parameterized distributional family $q$. The mean-field assumption uses an independent Gaussian distribution for each weight, and assumes the parameters $\theta = \{\mu, \sigma\}$ to be independent for each of the weights to make computation feasible. 

During training, the parameters $\theta$ are optimized towards minimizing the Kullback-Leibler divergence ($\mathrm{KL}$)~\cite{kullback1951information} between the variational distribution $q$ and the ideal weight distribution $p$.
This is encoded by the evidence lower bound ($\mathrm{ELBO}$)
\begin{align*}
    \mathrm{ELBO} = \underbrace{\mathbb{E}_{W\sim q}\left[\log p(Y|X,W)\right]}_{\text{data fitting}} - \underbrace{\lambda \, \mathrm{KL}\left(q||p\right)}_{\text{prior matching}} \quad 
\end{align*}
where $p(Y|X,W)$ is the probability of the training labels according to the model given the training data and weights, which are sampled from the variational distribution for the expected value, and $\lambda$ is an optional ``cold posterior'' hyperparameter factor~\cite{wenzel2020good}.

\autoref{fig:graphical-abstract} outlines how we take advantage of multiple parallelization axes within our implementation of \fancyname to circumvent the complexity and memory requirements inherent to the model. 

\subsection{Domain--Tensor Parallelism}
Our first optimization target is to reduce per-GPU memory footprint. 
With a single data item, i.e., atmospheric input state, being several hundred megabytes large, current state-of-the-art AI models for atmospheric forecasting struggle to go beyond the \num{1}-billion-parameter mark~\cite{wang2024orbit}. Accordingly, a significant proportion of the available GPU memory is consumed by data items and corresponding activations.
In the Bayesian formulation, this problem becomes even more pronounced, as the memory demand of a BNN compared to a deterministic model is increased proportionally to
\begin{enumerate*}
    \item the number of parameters used to parameterize the distribution of each model weight (two in our case); and
    \item the number of samples drawn---and thus, forward passes through the model and associated activations---per data item.
\end{enumerate*} 

Existing strategies for model parallelization are designed for training LLMs and thus prioritize maximizing the model size through tensor and pipeline parallelism. Long sequence lengths are handled by sharding the input data along the sequence dimension with context parallelism. Scientific image-based applications are much larger in terms of sequence lengths (order of \num{1} million) with the number of input variables in the order of \num{100}. The latent representations of the channels are projected to a larger embedding space; thus, sharding of the data along both the sequence and channel dimensions holds the biggest potential in alleviating memory bottlenecks.
We pursue this approach of \emph{domain parallelism} and combine it with tensor parallelism for all neural network layers.
That means that, in contrast to existing parallelization schemes, our approach distributes not only the model parameter and optimizer states, but also shards each input data item along both the spatial and channel dimensions. We leverage distributed matrix multiplications to implement this \emph{domain--tensor parallelization} (DTP) scheme~\cite{kieckhefen2026combining}. 
By sharding both the data and weights and formulating a linear layer as a distributed matrix multiplication, we are able to parallelize individual and arbitrary layers without the need for \texttt{allgather} operations or re-sharding, simultaneously maximizing GPU memory usage and reducing the data I/O-workload. 

We define a DTP instance over a GPU grid of [$c$ channels, $s$ spatial] ranks, where each rank loads $[1/c \times 1/s]$ fraction of each data item and holds $[1/c]$ of the model weights. Ranks within the same spatial group hold the same copy of the model weights, i.e., they compute on different data shards and are domain-parallel only. The parameters of the spatial groups are synchronized by \texttt{allreduce} operations of the gradients. Ranks within the same channel group are fully domain- and tensor-parallel. 

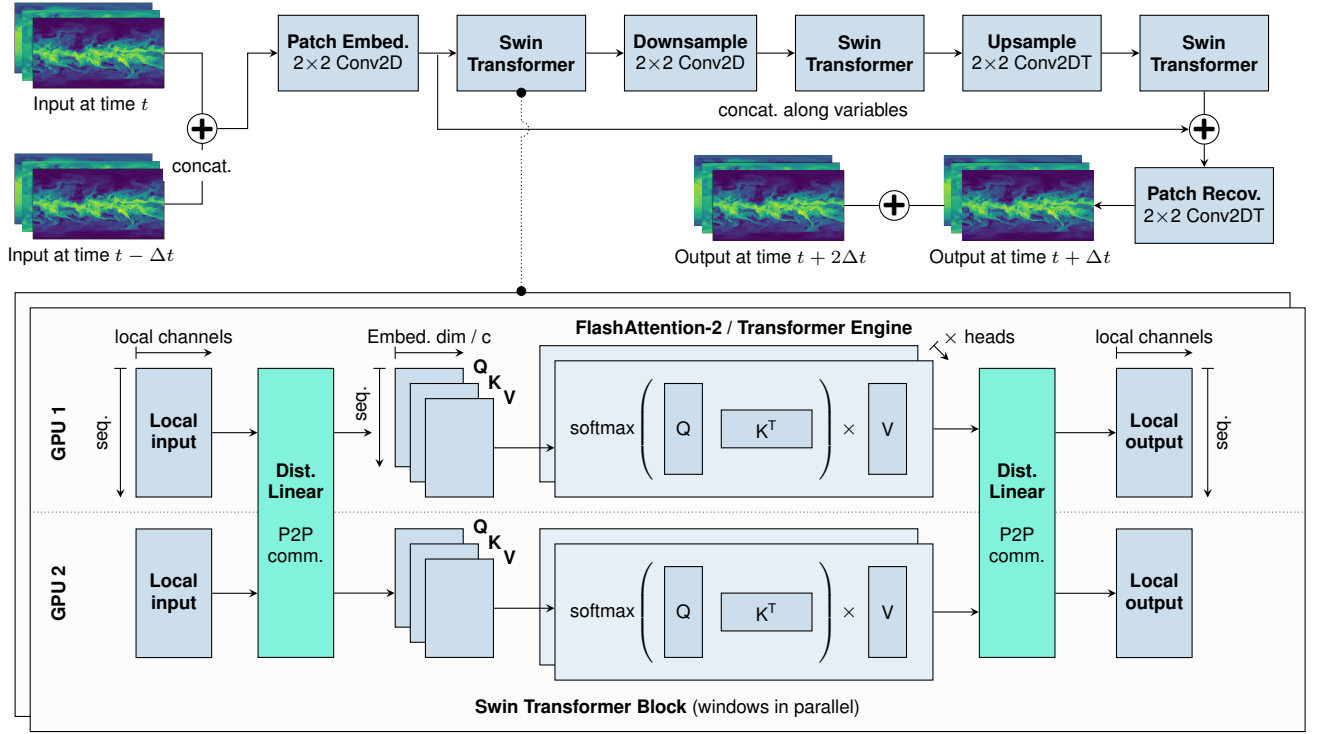
\begin{figure*}[ht!]
    \centering
    \input{figures/architecture.tikz}
    \caption{\fancyname combines a downsampling--upsampling backbone of Swin‑Transformer blocks with convolutional patch embedding and recovery. Unlike typical AI models, it ingests two consecutive samples ($t$ and $t-\Delta t$). The lower schematic illustrates the Swin blocks’ novel parallelization scheme using domain--tensor-parallelism.}
    \label{fig:architecture}
\end{figure*}

We apply this concept to all neural network layer types in \fancyname. Convolutions are performed by distributing the filters across the channel groups. Each rank computes the local convolutions and exchanges the necessary input data with the remaining ranks in the channel to resolve the full convolution. A schematic of the parallelization strategy used in channel-parallel attention layers is shown in \autoref{fig:architecture}.
The query--key--value (QKV) map in each attention layer is computed by a distributed linear layer across the channels. Each rank then holds the local sequence and a fraction of the QKV embedding dimension. The attention calculation is then trivially parallelized across the heads, allowing us to leverage optimized single GPU attention kernels such as FlashAttention-2~\cite{dao2024flashattention} or cuDNNs fused dot product attention via NVIDIA's Transformer Engine library~\cite{nvidia_transformer_engine} without any additional communication, while maintaining domain parallelism across channels on the input to the Transformer block. Head concatenation and the subsequent MLP are again performed by distributed layers, combining the information across the channel shards. The windows are shifted between every Transformer block by half a window, and halos are communicated between the ranks in spatial groups where necessary.

\subsection{Uncertainty Parallelism}
Next, we target the challenge of evaluating the $m$ forward passes of the random weight samples for each data item by introducing \emph{sampling parallelism}~\cite{ozdemir2026sampling}. 
We extend the GPU grid of [$c$ channels, $s$ spatial] ranks by an orthogonal parallelization axis with $u$ groups, along which we distribute the $m$ random weight samples. That means that we have $u$ DTP instances, each holding a full model on $c\times s$ GPUs. Each of these model instances draws $m/u$ independent random weight samples and performs the forward pass on them. 

To compute the mean and standard deviation of the ensemble, the sample predictions need to be aggregated: each rank computes the sum and squared sum of its local predictions, $S_u = \sum_{i=1}^{m_u} x_i^{(u)}$ and $Q_u = \sum_{i=1}^{m_u} \left(x_i^{(u)}\right)^2$, respectively. $S_u$ and $Q_u$ are then aggregated across all $u$ ranks in an \texttt{allreduce} operation to obtain the global sum and global square sum $S = \sum_{u} S_u,\quad Q = \sum_{u} Q_u$.
From these global sums, the global mean and global standard deviation of the predictions are computed.
This allows us to reduce the communication in the prediction aggregation to two tensors instead of all the $m$ predictions from the local weight samples.
Through sampling parallelism, we can efficiently increase the number of random weight samples per forward--backward pass, thereby improving prediction quality without being bound to the memory capacities of a single accelerator. 

Sampling parallelism provides an additional major advantage: the parallel evaluations of the random weight samples are all carried out on the same input data batch. They can therefore be combined with data augmentation, which improves predictive generalization and training convergence~\cite{brakel2024model}. While there are several data augmentation options from computer vision available, we chose a different route and used parallel data augmentation to include aleatoric uncertainty into our Bayesian model. 

We use a modification of conditional layer normalization~\cite{chen2021adaspeech}, inspired by \cite{alet2025skillful} (noise independent of spatial grid) and \cite{lang2026aifs}, which we coin \emph{noise input conditioning for ensembles} (NICE). In each forward pass, a random vector $\epsilon$ of $32$ independent standard normal distributed values is drawn. The \textit{scale} and \textit{shift} parameters for each layer normalization are generated by shallow MLPs, with the noise $\epsilon$ as the input. This effectively injects noise into the model and can itself be used to produce ensemble outputs, thus making the model fully uncertainty aware. We call this combination of sampling parallelism with random noise injection \emph{uncertainty parallelism} (UP). We handle the coordination of the data I/O within UP groups with synchronized random seeds.
Uncertainty parallelism allows us to evaluate more random weight samples by scaling to higher node counts, thereby improving the approximation of the current weight distribution and, ultimately, model quality.

\subsection{Data Parallelism}
Uncertainty parallelism can be combined orthogonally with data parallelism to accelerate and improve model training at the same time, adding an additional orthogonal scaling axis with $d$ groups to our GPU grid. Training with $d$ data parallel (DP) instances and $u$ uncertainty parallel instances thus requires $d\times u$ DTP instances in total. However, given the size of the input data items and the model, a realistic setup can only run with one or two local data items, i.e., local batch size, or random weight samples per GPU. Hence, data parallel scaling out also adds more data items to the effective global batch size.

In contrast to sampling parallelism, in which having larger sample numbers can lead to better predictive performance, scaling out data parallelism will lead to the occurrence of \emph{large batch effects}~\cite{keskar2017large}: AI models tend to converge to sharper local minima when the number of data items in a gradient update step becomes too large. For atmospheric models, this effect starts to arise at relatively small global batch sizes of around \num{16} to \num{32}~\cite{huber2026energy}. The degree to which model training can be accelerated through plain data parallelism is thus limited. With the additional usage of uncertainty parallelism, we circumvent this limitation. 

\subsection{Performance Optimization}
We further maximize per-GPU computational throughput with the following performance optimizations.
Since each GPU in a DTP instance loads a shard of the input data sample, we pre-shard the dataset accordingly in the channel and spatial (longitude) domain, thereby facilitating file access for data staging. To further stabilize data I/O and alleviate the strain on the file system, we perform data access in waves of \num{128} ranks in each wave and de-synchronize file system access every \num{50} iterations.

We fuse the MLPs of the NICE normalization and use grouped normalization to eliminate communication across the split channel domain.
Our implementation of the domain--tensor-parallel layers requires us to overwrite the backward function of each layer type. This allows us to also combine gradient communication in the backward pass with respect to the input and weights. This step essentially halves communication in the backward pass, significantly improving performance.

We handle communication for UP and DP groups via PyTorch's \texttt{DistributedDataParallel} wrapper~\cite{li2020pytorch}. We also apply post-grad accumulation hooks on the weight synchronization between spatial ranks to better overlap communication with the backward pass.

Additionally, we found that the computation of the ELBO loss takes significantly longer than the computation of a regular RMSE loss in a non-Bayesian neural network. To alleviate this bottleneck, we vectorize the computation in the prior matching term, which does not require communication, and overlap it with the computation for the data-fitting term, which does require communication of the locally obtained ensemble members.

\section{Performance Measurement}
We evaluate our 4D parallelization scheme, consisting of domain-, tensor-, uncertainty-, and data-parallelism, on the task of training \fancyname with mean-field variational inference.
\begin{table}[!t]
\sffamily
\centering
\caption{Software stack on JUPITER and LUMI}
\label{tab:software_stack}
\begin{tabularx}{\linewidth}{Xll}
\toprule
\textbf{Component}  & \textbf{JUPITER}  & \textbf{LUMI} \\
\midrule
Container Runtime   & Apptainer         & SingularityCE \\
Operating System    & RHE Linux 9.9     & SUSE 15.6 \\
Acc Library         & CUDA 13.2         & ROCM 6.2.4 \\
XCCL                & NCCL 2.29.7       & RCCL 2.22.5\\
PyTorch             & 2.12.0            & 2.7.1 \\
Python              & 3.12              & 3.12 \\
\bottomrule
\end{tabularx}
\vspace{-1em}
\end{table}
\subsection{Model Architecture}
The model architecture of \fancyname, visualized in~\autoref{fig:architecture}, is based on a 2D Swin Transformer~\cite{liu2021swin}: The input data is embedded via a 2D non-overlapping convolution with a spatial patch size of $2\times2$. Then, multi-headed self-attention is performed on non-overlapping windows in a Swin-Transformer-manner. In between each attention layer, the entire global domain is rolled by ($\lfloor \text{window size} / 2\rfloor$) patches in both spatial dimensions. Periodic boundaries are used in the latitude dimension. The data is then downsampled by a 2D convolution with a patch size of \num{2}, which also doubles the hidden dimension. Another Swin Transformer block is applied before the data is upsampled back with a transposed convolution and an MLP to mix across the channels, before another Swin Transformer block is applied. Finally, the target, with the original spatial dimensions, is recovered with a transpose convolution.

\subsection{Training objective}
We train \fancyname to perform \num{12}- and \num{24}-hourly forecasts based on two previous time steps, i.e., atmospheric variables at times $t$ and $ t-12 h$ are used as input to predict atmospheric variables at times $(t + 12h, t+24h)$.
For training, we use the ERA5 dataset~\cite{hersbach2020era5}, obtained from WeatherBench2~\cite{rasp2024weather} and stored in the zarr format. The data uses an equiangular grid at a global resolution of \ang{0.25} ($\approx$ \qty{30}{\kilo\meter}). We consider \num{84} input variables in total---six surface variables: \qty{10}{\meter} u-velocity, \qty{10}{\meter} v-velocity, mean sea level pressure, \qty{2}{\meter} temperature, sea surface temperature, total precipitation, and six pressure variables: geopotential, specific humidity, temperature, and the u, v, and w components of wind velocity. Pressure variables are considered at \num{13} pressure levels of \qtylist[list-final-separator = {, and },list-units = single]{1000;925;850;700;600;500;400;300;250;200;150;100;50}{\hecto\pascal}. Constant inputs of soil type, topography, and land masks are also used. Variables from different time steps are concatenated along the channel dimension. Data is normalized with per-variable Z-score normalization. 
Hourly data from 1979--2019 is used for training, and data from 2020--2025 is used for evaluations. 

For minimizing the ELBO as loss function, we use an Adam optimizer with an initial learning rate of $10^{-4}$, decaying to $10^{-5}$ with a cosine decay. Linear warm-up is used for the first \num{10000} gradient steps. 
We train the model using mixed precision format: for Transformer Engine and FlashAttention in the Transformer layers, we leverage Bfloat16 precision. All other computations are performed in TensorFloat32 or Float32 precision, to reflect the sensitivity of the Bayesian optimization objective to numerical rounding errors.

\subsection{System specifications}
We perform training of \fancyname on two systems: JUPITER and LUMI.

JUPITER is a EuroHPC JU supercomputer hosted at Forschungszentrum Jülich, Germany. It is the first exascale machine in Europe and ranks \#5 in the TOP500 of June 2026, reaching \qty{1}{\exa \flops} in the HPL benchmark. JUPITER has been designed for energy-efficient, high-scaling numerical and AI applications~\cite{herten2024application}, and is currently in the finalization of its build-up phase.
The system utilizes \num{23536} NVIDIA GH200 superchips, each combining a \num{72}-core NVIDIA Grace CPU (\qty{120}{\giga\byte} LPDDR5 memory) with a Hopper GPU (\qty{96}{\giga\byte} HBM3 memory) through a unique high-bandwidth bus called NVLink-C2C (\qty{900}{\giga\byte\per\second} bandwidth). The full system features \num{5884} nodes, of which about 5200 were accessible to us. Within a node, the GPUs are all-to-all connected with pair-wise bandwidths of \qty{300}{\giga\byte\per\second}. Similarly, all CPUs are all-to-all-connected with pair-wise bandwidths of \qty{200}{\giga\byte\per\second}. 

Each node features four NVIDIA InfiniBand NDR200 network adapters with each \qty{200}{\giga\bit\per\second} injection bandwidth, interconnected in a Dragonfly+ topology. Up to \num{240} nodes are combined in a Dragonfly group, internally connected over two levels of switches in a fat tree (local) topology. Twenty-five of these groups form the JUPITER network, with \num{30} direct links between each pair of groups; each inter-switch link provides \qty{400}{\giga\bit\per\second} bandwidth. Adaptive routing is enabled to counteract congestion due to the tapering of the network outside of the groups.

LUMI is a EuroHPC JU supercomputer hosted by the CSC – IT Center for Science in Finland. It is a pre-exascale Cray EX supercomputer supplied by Hewlett Packard Enterprise (HPE). As of June 2026, LUMI ranks \#11 in the TOP500 list, reaching \qty{386}{\peta \flops} in the HPL benchmark.
The majority of the system's compute power is found in the LUMI-G hardware partition, which consists of \num{2978} nodes. Each node is equipped with four AMD MI250x GPUs and a single \num{64}-core AMD EPYC Trento CPU. The MI250X GPU is a multi-chip module containing two GPU dies named Graphics Compute Die (GCD) by AMD. Each die includes \num{110} compute units and has access to a \qty{64}{\giga\byte} slice of HBM2e memory, resulting in a total of \num{220} compute units and \qty{128}{\giga\byte} of total memory per MI250x module.

LUMI uses a Dragonfly network topology in which compute nodes are organized into groups interconnected via switches. The compute nodes are linked through the HPE Cray Slingshot-11 \qty{200}{\giga\bit\per\second} network interconnect (NIC). The HPE Cray Slingshot NIC features high-performance RDMA and hardware acceleration for MPI and SHMEM-based software.

\begin{table*}[ht!]
\sffamily
\small
\centering
\caption{\fancyname configurations used in scaling experiments and model training.}
\label{tab:model_configuration}
\begin{tabularx}{\linewidth}{Xcrcccccccc}
\toprule
& & &\multicolumn{4}{c}{\textbf{Outer windows}} & \multicolumn{4}{c}{\textbf{Inner windows}} \\ 
\cmidrule(lr){4-7} \cmidrule(lr){8-11} 
Name  & Params (B) & \unit{\tera\flop} & Dim. & Win. size & Heads & Layers & Dim. & Win. size & Heads&  Layers  \\
\midrule
\largejup & \num{2.36} & 3557.9&\num{3072} & [\num{360}, \num{180}] & \num{24} & \num{2} & \num{6144} & [\num{180}, \num{30}] & \num{48} & \num{2}\\
\largelumi & \num{1.05} & 2238.2 &\num{2048} & [\num{360}, \num{180}] & \num{16} & \num{2} & \num{4096} & [\num{180}, \num{30}] & \num{32} & \num{2}\\
\smalljup & \num{0.68} & 423.8 &\num{1024} & [\num{30}, \num{60}] & \num{16} & \num{4}  & \num{2048} & [\num{20}, \num{30}] & \num{32} & \num{6}\\
\bottomrule
\end{tabularx}
\vspace{-1em}
\end{table*}

\subsection{Software}
The software stack employed in this work is based on containerized environments, leveraging Apptainer / Singularity Community Edition~\cite{kurtzer2017singularity} to ensure portability and reproducibility across high-performance computing (HPC) systems.
Apptainer enables unprivileged container execution and is widely adopted in scientific and high-performance computing contexts.
On JUPITER, the base container is the NVIDIA NGC PyTorch image (version \texttt{26.04-py3}~\cite{nvidia2026pytorch}), available on the NVIDIA NGC registry. For LUMI, we rely on a provided custom ROCm+PyTorch container~\cite{sfantao2026lumi}.

Domain and tensor parallelism of the PyTorch layers was done by using the library \texttt{jigsaw}~\cite{kieckhefen2026combining}, and the Bayesian model was obtained by converting the deterministic torch model with \texttt{torch\_blue}~\cite{weyrauch2026torch_blue}. Profiles and traces for optimization were recorded using the built-in \texttt{torch} profiler and NVIDIA Nsight Systems 2026.2.1. Power draw and memory utilization measurements for the hardware accelerators were sampled during training using \texttt{perun} \cite{gutierrez2023perun} for both systems. 

\subsection{Performance metrics}
The most relevant performance metrics when training neural networks are predictive performance and overall training time. For a fixed-size dataset, training time is influenced by two factors: the time it takes to iterate through the entire dataset in batched forward-backward passes (\emph{time-per-epoch}) and the number of epochs, i.e., iterations over the entire dataset, necessary for model convergence (\emph{convergence rate}). While the latter is generally addressed algorithmically, the former can be directly tied to the computational throughput of performing one batch iteration (forward--backward pass and weight update). 

To assess the computational throughput of the batch iterations in our scaling experiments, we measure the time for 
\begin{enumerate*}
    \item a forward--backward-update step, and 
    \item one full epoch 
\end{enumerate*}
using Python's high-resolution \texttt{time} module with \texttt{time.perf\_counter\_ns()} and the \texttt{torch.cuda.streams.Event()} to set markers to monitor accelerator computations.
The former includes the overhead associated with model training: data I/O, checkpointing, and evaluation; whereas the latter captures exclusively the computation that is offloaded to hardware accelerators.
We further estimate the number of floating-point operations (\unit{\flop}) of a  forward--backward pass on a DTP instance, using PyTorch's \texttt{torch.utils.flop\_counter} tools.
Based on this, we report the number of \unit{\flops} for 
\begin{enumerate*} 
    \item a forward--backward-update-step as \emph{peak performance} and 
    \item for a full epoch as \emph{sustained performance} 
\end{enumerate*} by dividing the number of \unit{\flop} by the measured time. We also report accelerator metrics related to ideal use during the forward--backward-pass, including percentage of idle time, memory utilization, and average power draw.

In addition to these computational throughput numbers, we also report the commonly used metric \emph{images/sec} from computer vision.  In atmospheric modeling, this corresponds to the \emph{forecast throughput} or \unit{\forecastpsec}, which is estimated based on the average batch processing time, and the total number of data items per batch (DP instances $\times$ local batch size). We also introduce the orthogonal metric of \unit{\samplepsec}, determined via the product of \unit{\forecastpsec} $\times$ UP instances $\times$ local sampling size, for \emph{ensemble generation throughput}. This metric indicates throughput regarding the evaluation of random weight samples of the BNN and is especially relevant for domain scientists during inference over an evaluation dataset, indicating how long it takes to generate a meaningful ensemble.

Next to compute performance, we evaluate the probabilistic predictive skill scores of a fully trained \fancyname, where we report the commonly used root mean square error (RMSE) of the ensemble average prediction, probabilistic spread skill ratio (SSR), and continuous ranked probability score (CRPS) metrics for different atmospheric variables. 
These metrics cover three different aspects: The RMSE of the ensemble mean measures how well the average forecast matches the target. The SSR quantifies whether the model's uncertainty is well-calibrated. The CRPS, which we calculate using the fair CRPS estimator~\cite{zamo2018estimation}, combines these two aspects to measure overall probabilistic prediction quality.
We compare our predictive skill scores against the current-best probabilistic AI model, FGN~\cite{alet2025skillful} vs. ERA5, and the IFS ensemble forecast~\cite{ecmwf2024ifs} vs. ERA5, both obtained from~\cite{rasp2024weather}. 

We further assess the skill of \fancyname~on predicting extreme events, by evaluating three exemplary use-cases, using the recently published ExtremeWeatherBench (EWB) Python package~\cite{mcgovern2026extreme}:
\begin{enumerate*}[label=\alph*)]
    \item a heat wave,
    \item a freeze, and
    \item a tropical cyclone (TC). 
\end{enumerate*}
The heat wave occurred from August 8, 2022, until August 16, 2022, and affected southern UK, France, and Spain. The freeze affected Western and Central Europe from December 10, 2022, until December 20, 2022.
For the TC, we used Damien, a category 3 cyclone, which occurred between February 3, 2020, and February  9, 2020, over Western Australia. These cases probe both prolonged large-scale temperature anomalies and the emergence and subsequent trajectory of a compact, rapidly evolving severe-weather system.

\section{Performance Results}
\autoref{tab:model_configuration} describes the three model configurations used in our experiments. 
The \largejup~and \largelumi~model architectures are optimized to leverage the available hardware and train the maximal model size as efficiently as possible. 
Due to the size of these model configurations, the local batch size and number of local samples are restricted to \num{1}. However, the long sequence length considered in the attention windows (\num{64000}) causes the training time per sample to become prohibitively long for training a model until convergence. Therefore, we define~\smalljup, a smaller but deeper model configuration designed to maximize predictive skill.

For all three configurations, we distribute one DTP instance on a GPU mesh of $[s,c]=[4,2]$, i.e., \num{8} GPUs (\num{2} nodes) on JUPITER and \num{8} GCDs (\num{1} node) on LUMI, when performing training steps (forward--backward-update passes). At inference, where the memory footprint is much smaller due to the absence of gradients and optimizer states, we distribute one DTP instance on a GPU mesh of $[s,c]=[2,2]$, i.e., \num{4} GPUs.

\subsection{Baseline performance}
We define baseline performance on the \largelumi~and \largejup~model configurations. 
The baseline configurations require \num{16} GPUs on both systems: a domain--tensor group of [\num{4}, \num{2}] $\times$ \num{2} UP groups, and a single data parallel group. This translates to \num{4} compute nodes on JUPITER, and \num{2} nodes on LUMI.

Based on an average batch time (forward--backward-update) of \qty{1.68}{\second} on the JUPITER system, and an estimated \qty{3.56}{\peta\flop} operations given the model configuration, \largejup~reaches a peak performance of \qty{1.06}{\peta\flopsnode}, and a sustained performance of \qty{0.92}{\peta\flopsnode}. This translates to \qty{0.59}{\forecastpsec}. We measured the amount of idle time and time spent waiting for communication by profiling the system on \num{20} gradient updates, showing $\sim$\qty{25}{\percent} of the time is spent waiting on communication, $\sim$\qty{1}{\percent} device idle time, and $\sim$\qty{72}{\percent} of the time is being used for computation. Within an epoch, memory measurements for the individual GPUs report \qty{71.45}{\giga\byte} consistently, utilizing \qty{74.43}{\percent} of the total accelerator memory. 
With the \largelumi~configuration, the average forward--backward-pass runs for \qty{4.47}{\second}. Based on an estimated \qty{2.24}{\peta\flop}, we reach single node peak performance of \qty{0.5}{\peta\flopsnode}, or \qty{0.22}{\forecastpsec}. Memory measurements show each GCD reporting an average of \qty{48.87}{\giga\byte}, or \qty{76}{\percent} GPU memory utilization. 

Additionally, we measured the power draw of the hardware accelerators during training to provide a realistic estimate of the energy cost of training and running our Bayesian model~\cite{debus2023reporting}. We observe an average power draw of \qty{455.6}{\watt} on each MI250X card, and \qty{3.64}{\kilo\watt} for training~\largelumi~in the baseline configuration. By comparison, we measured a consistent power draw of \qty{514.36}{\watt} per GPU on the JUPITER system, resulting in a total GPU power draw of \qty{8.22}{\kilo\watt} for training~\largejup~over the four nodes in the baseline configuration.

\begin{figure*}[ht!]
    \centering
    \sffamily
    \ref*{leg:hero-legend}\quad\ref*{leg:sample-legend}
    \captionsetup[subfigure]{justification=centering}
    \begin{subfigure}[t]{0.48\linewidth}
        \caption{Weak scaling computational throughput}
        \input{plots/throughput-scaling.tikz}
    \end{subfigure}
    \begin{subfigure}[t]{0.48\linewidth}
        \caption{Strong scaling performance}
        \input{plots/dp-scaling.tikz}
    \end{subfigure}
    \begin{subfigure}[t]{0.48\linewidth}
        \caption{Weak scaling ensemble throughput}
        \input{plots/sample-scaling.tikz}
    \end{subfigure}
    \begin{subfigure}[t]{0.48\linewidth}
        \caption{Strong scaling data throughput}
        \input{plots/forecast-scaling.tikz}
    \end{subfigure}
    \caption{Scaling results of \fancyname on JUPITER and LUMI for various data‑ and uncertainty-parallel setups, expressed in computational, data, and ensemble throughput as well as peak performance under strong scaling, i.e., increasing DP with increasing node counts, and weak scaling with fixed per-node workload, i.e., increasing UP with increasing full GPU counts.}
    \label{fig:scaling-experiments}
\end{figure*}
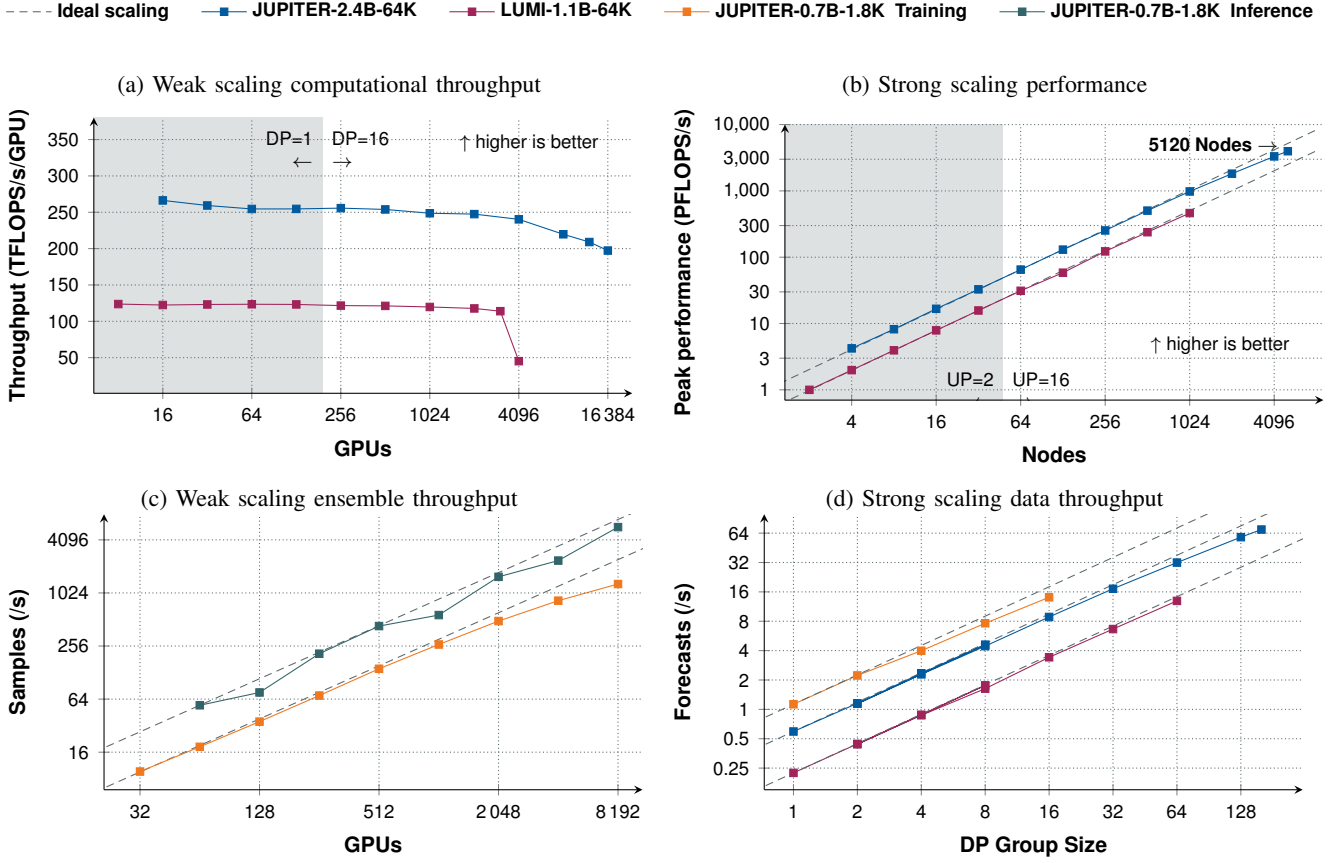
\subsection{Scaling experiments}
To evaluate the efficiency of uncertainty parallelism, we scale out the number $u$ of UP groups in a weak scaling manner, i.e., we increase the number of global weight samples evaluated per data batch by increasing the number of UP instances. For~\largejup~and~\largelumi~we conduct this experiment in two regimes: for one, we scale $u$ from \num{2} to \num{16}, with a single DP instance, i.e., using \num{16} through \num{128} GPUs, or GCDs in LUMI. Such a configuration is not very realistic for production training runs, where multiple DP instances would be used to iterate over the dataset faster. Hence, we conduct a second round of this scaling experiment, where we keep the number of DP instances, i.e., the global batch size, fixed at \num{16} and increase the number of UP instances from \num{2} to \num{64}, i.e., using \num{256} to \num{16384} GPUs. 
For~\smalljup~we evaluate UP scaling in both training and inference mode, with a fixed global batch size of \num{4}, increasing the number of UP instances from \num{8} to a total of \num{2048} nodes on JUPITER. 
The scaling experiments shed light on two interesting aspects of our multidimensional parallelization scheme. For one, DP requires different data to be read from storage by each rank, while for UP, each rank reads the same data. Having both UP and DP instances participate in the training illustrates how our parallelization scheme handles these different I/O streams.
Second, both UP and DP utilize PyTorch's \texttt{DistributedDataParallel} wrapper~\cite{li2020pytorch} to handle communication. Thus, the experiment allows us to investigate if these two communication axes potentially compete against one another, thereby hindering performance.

The results for~\largejup~and~\largelumi~are shown in \autoref{fig:scaling-experiments} (a). The values in the gray section indicate the scaling experiments where the DP size is \num{1}, and the values after the gray section showcase the results of the configuration with the increased DP size of \num{16}.
On LUMI, we observe almost constant throughput for lower node counts up to \num{3072} GPUs, whereas on JUPITER, we observe a small degradation of per-GPU performance. On both systems, performance starts to deteriorate notably at larger node counts. This scalability limit of UP can be attributed to the increased stress on the file system and higher communication cost during the loss calculation.
Ensemble throughput scaling of~\smalljup~is presented in \autoref{fig:scaling-experiments} (c).
On \num{2048} nodes, the training configuration reaches a total of \qty{1293.31}{\samplepsec}, with a scaling efficiency of \qty{52.35}{\percent}, similar to the results shown in \autoref{fig:scaling-experiments} (a). Again, we observe a degradation in performance, which can be attributed to the additional communication required to calculate the loss for a larger number of samples generated per forward pass. 
This effect disappears in inference mode, where the loss is not calculated, and the uncertainty parallel groups become independent of each other. On \num{2048}, the predictive model reaches \qty{5716.61}{\samplepsec}, with an ensemble scaling efficiency of \qty{82.01}{\percent}. 

Next, we evaluate scaling out data parallelism in a strong scaling manner: We increase the global batch size by increasing the number of DP instances from \num{1} to \num{16} at a fixed local batch size of \num{1}, thereby reducing the overall number of iterations in an epoch. Similarly to UP weak scaling, we conduct this strong scaling experiment for~\largejup~and~\largelumi~in two setups: one with just \num{2} UP instances, thus scaling from \num{2} to \num{32} nodes, and one with \num{16} UP instances, thus scaling from \num{64} to \num{5120} nodes.
For~\smalljup~we keep the number of random weight samples (ensemble members) in the training configuration at \num{96}, using \num{48} UP instances, and scale out DP from \num{1} to \num{16}, i.e. \num{96} to \num{1536} nodes. Results are presented in \autoref{fig:scaling-experiments} (b) and (d). Again, the values in the gray sections indicate the scaling experiments where the UP size is \num{2}, i.e., training with a \num{2} UP instance, and the values after the gray section showcase the results of the configuration with the increased UP size of \num{16}.


We observe near-perfect DP-strong scaling for all three model configurations on both JUPITER and LUMI. On \num{5120} nodes on JUPITER, i.e., \num{20480} GH200 GPUs,~\largejup~runs at a total peak performance of \qty{3.96}{\exa\flops} during forward--backward-update step, and a scaling efficiency of \qty{73.99}{\percent}. \largelumi~ reaches \qty{463.23}{\peta\flops} at \num{1024} nodes, corresponding to \qty{92.09}{\percent} scaling efficiency. These results indicate that the parallelism strategy and model configuration space allow for good system portability. The seamless transition between the different UP and DP instance sizes highlights the efficiency of both parallelism axes. 

\smalljup~yields a higher forecast throughput per data parallel group than the large configurations. 
At a single DP instance, the predictive configuration achieves a \qty{1.13}{\forecastpsec}, almost double the throughput of~\largejup. 
With 16 DP instances, or \num{1536} nodes, it reaches \qty{14.08}{\forecastpsec}, each with \num{96} ensemble members, at a sustained performance of \qty{239.88}{\tera\flop}.

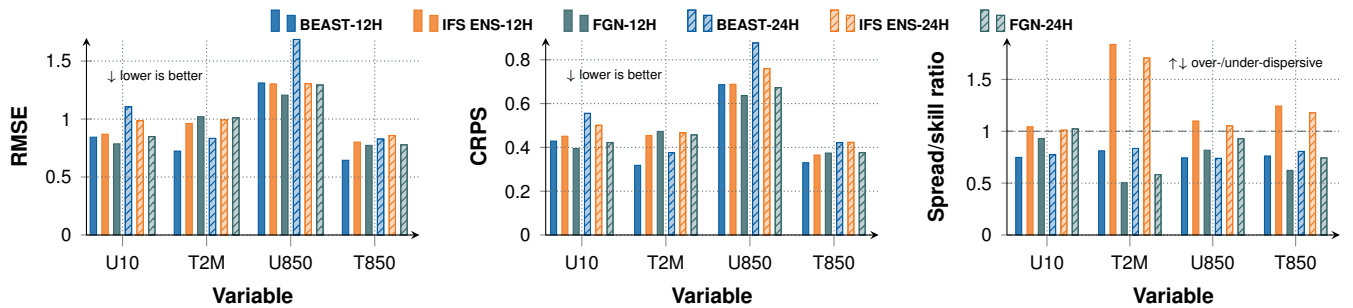
\begin{figure*}[ht!]
    \centering
    \sffamily
    \ref*{leg:predictive-performance-legend}
    
    \begin{subfigure}{0.33\linewidth}
        \centering
        \input{plots/rmse_new.tikz}
    \end{subfigure}
    \begin{subfigure}{0.33\linewidth}
        \centering
        \input{plots/crps_new.tikz}
    \end{subfigure}
    \begin{subfigure}{0.33\linewidth}
        \centering
        \input{plots/spread-skill-ratio_new.tikz}
    \end{subfigure}
    \caption{Forecasting skill of \fancyname across the four key variables, $u$ component of wind velocity at 10m (U10), 2 meter temperature (T2M), $u$ component of wind velocity at \qty{850}{\hecto\pascal} (U850), and temperature at \qty{850}{\hecto\pascal} (T850), compared to IFS-ENS and FGN. All three models were evaluated against ERA5, on one year of data (2022).}
    \label{fig:predictive-performance}
\end{figure*}

\subsection{Full model training and predictive skill}
We train \fancyname~in the~\smalljup~configuration on \num{384} nodes on JUPITER, using \num{4} DP and \num{48} UP instances, for \num{983350} optimizer update steps.
We use \num{2} samples per UP instance and keep the local batch size to \num{1}, thus using a global batch size of \num{4} and drawing \num{96} random weight samples per optimizer update step.
One epoch on 40 years of data (1979--2020) requires \num{89835} gradient update steps and takes \qty{25.45}{\hour}. The sustained training performance is \qty{70.46}{\peta\flops}.
Predictive metrics of the trained model for four different variables (U10, T2M, U850, T850) for the two forecast time steps \qty{12}{\hour} and \qty{24}{h} are shown in \autoref{fig:predictive-performance}, in comparison to the state-of-the-art probabilistic model FGN and the IFS ensemble forecast. We observe competitive to superior predictive scores with much better calibrated uncertainty estimates, and stable roll-out for up to ten days.

\begin{figure}[ht!]
    \centering  
    \includegraphics[width=0.95\linewidth]{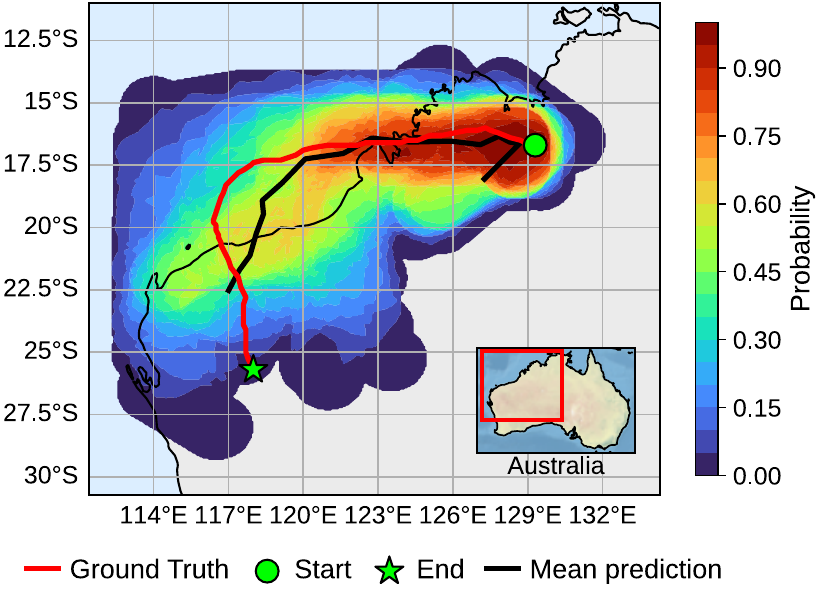}
    \caption{Probability for cyclone passing within \qty{150}{\kilo\meter} for TC Damien, predicted by \fancyname~(initialized with ERA5 data from 02-01-2020, 96 ensemble members). The reference track is obtained from the International Best Track Archive for Climate Stewardship (IBTrACS).}
    \label{fig:tc-damien}
    \bigskip
    \begin{subfigure}{\linewidth}
    \input{plots/rmse_comparison_heat_wave.tikz}
    \end{subfigure}
    \begin{subfigure}{\linewidth}
    \input{plots/rmse_comparison_freeze.tikz}
    \end{subfigure}
    \caption{RMSE values of \fancyname~predictions for a heat wave and a freeze, both occurring over Europe in 2022, compared to the IFS HRES. Predictions were generated with \num{32},\num{96} and \num{256} ensemble members, rolled out for up to 10-days, with \qty{6}{\hour}- initialization times. Lower values indicate better predictive skill.}
    \label{fig:heat-wave-freeze}
\end{figure} 

Furthermore,~\fancyname~can predict extreme events with exceptional skill. For one, it excels at the task of TC track prediction, as shown in \autoref{fig:tc-damien} for TC Damien. The mean ensemble prediction of \fancyname~follows the observed track, including the transition from land to ocean and the subsequent landfall. The ensemble members form a spread that encloses the observed track, with uncertainty increasing with forecast lead time as expected. Second, it can forecast both the heat wave and the freeze several days in advance, with skill scores competitive to or even superior to the IFS, as shown in \autoref{fig:heat-wave-freeze}. These results show that \fancyname's probabilistic forecasts remain informative not only in aggregate verification, but also for rare events where accurately representing both the expected evolution and its uncertainty is particularly consequential.

\noindent Increasing the number of random weight samples not only impacts computational performance but also predictive skill. \autoref{tab:ensemble-generation-performance} illustrates how larger ensemble sizes lead to better predictive skill in terms of lower RMSE. This result highlights the benefit of our uncertainty parallelism approach, which allows us to efficiently generate large ensembles by scaling out uncertainty parallelism. Generating a 15-day forecast trajectory with \fancyname~takes about \qty{16}{\second} to \qty{17}{\second} on \num{4} GH200 GPUs, independent of the ensemble size. For comparison, FGN reports inference times of just under 1 minute on a single TPU v5p~\cite{alet2025skillful}. 

In summary, our results demonstrate the strengths of our innovations: They enable us to \textbf{efficiently utilize} the available resources (up to \num{20480} GH200 GPUs) to generate \textbf{large, accurate ensembles}, while allowing for \textbf{high flexibility} regarding the modeling objectives. The combination of domain-, tensor-, uncertainty, and data-parallelism in an orthogonal 4D parallelization scheme can be leveraged to: 
\begin{itemize}[noitemsep]
    \item prioritize model size beyond the 1B parameter mark towards building \textbf{bigger} models, e.g., foundation models 
    \item prioritize weight sample number towards maximizing ensemble generation throughput for \textbf{better} uncertainty quantification
    \item prioritize batch size towards maximizing forecast throughput for \textbf{faster} training and inference 
\end{itemize} 
Together, these aspects allow us to leverage the full potential of the Bayesian paradigm for \textbf{smarter} AI models in atmospheric forecasting.

\begin{table}[ht!]
\sffamily
\small
\centering
\caption{Inference time for a 15-day forecast, and RMSE values of 2-meter temperature (T2m) for \qty{12}{\hour} and \qty{24}{\hour} predictions generated with~\fancyname-\smalljup, evaluated on one year of validation data (2022) with increasingly larger ensembles. Evaluation was conducted by distributing one DTP instance on \num{4} GH200 GPUs, and a global batch size of 16.}
\label{tab:ensemble-generation-performance}
\begin{tabularx}{\linewidth}{r>{\centering\arraybackslash} p{1cm}>{\centering\arraybackslash}X>{\centering\arraybackslash}cc}
\toprule
\multirow{2}{*}{\# GPUs} & \multirow{2}{=}{\centering Ensemble member} & \multirow{2}{=}{\centering Inference time [s]}&\multicolumn{2}{c}{T2m RMSE}\\ 
\cmidrule(lr){4-5}
&   &  & 12h  & 24h  \\ 
\midrule
128 & 8 & 15.58 & 0.7507 & 0.8661  \\
512 & 32 & 16.85 & 0.7289 & 0.8390  \\
1536 & 96  & 16.98 & 0.7237 & 0.8328 \\
3072 & 768 & 15.74 & 0.7226 & 0.8316  \\ 
\bottomrule
\end{tabularx}
\vspace{-1em}
\end{table}

\section{Implications}
Until now, training AI models for global atmospheric forecasting in a fully Bayesian framework has not been possible. While implementing the complex mathematical concepts behind BNNs is already a challenge in itself, the associated computational training cost has ultimately been the prohibitive factor. We present \fancyname, the first large-scale Bayesian Swin Transformer for atmospheric forecasting, able to quantify both, epistemic (model) and aleatoric (data) uncertainty. This breakthrough is enabled by our innovative 4D parallelization scheme, consisting of domain-, tensor-, uncertainty-, and data-parallelism.

We achieve state-of-the-art performance in both training speed and efficiency, peaking at \qty{3.96}{\exa\flops} on \num{20480} NVIDIA GH200 GPUs on the JUPITER supercomputer. 
After training for nearly one million gradient updates, a \num{700}-million-parameter model achieves predictive skill scores competitive with the state-of-the-art probabilistic AI model FGN and the IFS ensemble forecast. Yet, through our highly efficient domain-tensor-parallelism approach, \fancyname~is able to generate large ensembles significantly faster: a single 15-day forecast can be generated on \num{4} GH200 GPUs in less than \qty{20}{\second}.
On \num{1536} nodes, we achieve a forecast generation throughput of \qty{14.08}{\forecastpsec} with 96 ensemble members.  

This innovation opens a new era for AI-based atmospheric sciences. Under a changing climate, fast and accurate uncertainty quantification in weather forecasting is becoming increasingly important. Existing approaches to uncertainty quantification in AI models follow methods from generative AI, using input randomness to produce ensemble forecasts; however, they neglect the uncertainty of the model itself, making the predictions overconfident~\cite{nath2026can}. In contrast, our Bayesian model formulation allows for a more principled estimation of the prediction certainty, leading to better calibrated uncertainty estimates. 
This calibration is crucial, especially for predicting extreme weather events such as heatwaves, floods, and severe storms. Decision makers depend on accurate estimates of the likelihood and severity of extreme events to prepare for possible outcomes and take preventative measures to reduce potential damages to people, infrastructures, and the economy. As demonstrated by our results, \fancyname~can deliver upon this promise by predicting extreme events, such as heat waves or tropical cyclones, with high accuracy.
Our contribution thus directly addresses the challenges we are facing under a changing climate.

Next to these direct implications, our approach to enable fast and efficient training of large-scale BNNs also holds the potential to revolutionize Earth system sciences beyond atmospheric forecasting.
While the lack of sufficiently large, well-curated datasets has thus far hindered true breakthroughs in AI-based models of Earth system components other than the atmosphere---such as the ocean---our Bayesian approach can overcome this barrier. With its capabilities to learn the underlying data distribution and to capture the associated uncertainty in its distributional model weights, a BNN is particularly well-suited to leverage small, high-quality datasets, e.g., from km-scale climate simulations~\cite{klocke2025computing} to build the next generation of climate emulators. Our approach to flexibly allocate GPU memory through highly efficient domain--tensor-parallelism puts us in a prime position to increase the size of input data items beyond the current resolutions towards multi-timestep forecasting and km-scale modeling.

Our breakthroughs were enabled by leveraging top-tier supercomputers with thousands of GPUs. Going forward, the results presented in this work mark only the beginning of exploring high-performance computing for BNNs. By introducing uncertainty parallelism as an orthogonal scaling axis, we have successfully overcome the current barriers in scaling out neural network training through plain data-parallelism, which is inevitably limited due to large-batch effects. Scaling out uncertainty parallelism goes beyond merely accelerating neural network training---it creates scientific value.
For future work, we aim to leverage the power of current and future exascale systems by improving stability for lower precision computations and seeking out additional parallelization axes for scaling.

\section*{Acknowledgment}
    This project received access to the JUPITER supercomputer through the JUPITER Research and Early Access Program (JUREAP). JUPITER is funded by the EuroHPC Joint Undertaking, the German Federal Ministry of Research, Technology and Space, and the Ministry of Culture and Science of the German state of North Rhine-Westphalia.
    Computations on the LUMI supercomputer were enabled by resources provided by CSC - IT Center for Science, Finland. LUMI is co-funded by the EuroHPC Joint Undertaking and the LUMI Consortium.
    
    We would like to express our gratitude to Christelle Piechurski from NVIDIA for the continuous support and encouragement to pursue this project from the beginning. We would like to thank Benedikt von St. Vieth and Damian Alvarez from JSC for their assistance and support, and Christian Schiffer for the exchanges on extreme-scale AI. We are also thankful to Jesse Harrison and Jarmo Mäkelä from CSC for supporting this submission, and to Gaurav Naithani and Billy Braithwaite from CSC for their continuous input and insightful discussions on the topics of Bayesian neural networks and FlashAttention.  

    This research is supported by the German Federal Ministry of Research, Technology and Space (BMFTR) under the 01LK2313A SMARTWEATHER21-SCC-2 grant, by the Carl Zeiss Foundation under the Breakthroughs 2025 ``WOW – a World model of Our World'' project, by the European High Performance Computing Joint Undertaking (JU) and the the German Federal Ministry of Research, Technology and Space (BMFTR), the Ministry of Culture and Science of North Rhine-Westphalia (MKW NRW) and the Hessian Ministry of Science and Research, Arts and Culture (HMWK) under grant agreement No 101250682 (JUPITER AI Factory, JAIF).
    
\bibliographystyle{IEEEtran}
\bibliography{references}
\end{document}

%% file: colors.tex
\definecolor{hgfblue}{RGB}{0, 90, 160}
\colorlet{hgfblue10}{hgfblue!10!white}
\colorlet{hgfblue20}{hgfblue!20!white}
\colorlet{hgfblue30}{hgfblue!30!white}
\colorlet{hgfblue40}{hgfblue!40!white}
\colorlet{hgfblue50}{hgfblue!50!white}
\colorlet{hgfblue60}{hgfblue!60!white}
\colorlet{hgfblue70}{hgfblue!70!white}
\colorlet{hgfblue80}{hgfblue!80!white}
\colorlet{hgfblue90}{hgfblue!90!white}

\definecolor{hgfdarkblue}{RGB}{0, 40, 100}
\colorlet{hgfdarkblue10}{hgfdarkblue!10!white}
\colorlet{hgfdarkblue20}{hgfdarkblue!20!white}
\colorlet{hgfdarkblue30}{hgfdarkblue!30!white}
\colorlet{hgfdarkblue40}{hgfdarkblue!40!white}
\colorlet{hgfdarkblue50}{hgfdarkblue!50!white}
\colorlet{hgfdarkblue60}{hgfdarkblue!60!white}
\colorlet{hgfdarkblue70}{hgfdarkblue!70!white}
\colorlet{hgfdarkblue80}{hgfdarkblue!80!white}
\colorlet{hgfdarkblue90}{hgfdarkblue!90!white}

\definecolor{hgflightblue}{RGB}{20, 200, 255}
\colorlet{hgflightblue10}{hgflightblue!10!white}
\colorlet{hgflightblue20}{hgflightblue!20!white}
\colorlet{hgflightblue30}{hgflightblue!30!white}
\colorlet{hgflightblue40}{hgflightblue!40!white}
\colorlet{hgflightblue50}{hgflightblue!50!white}
\colorlet{hgflightblue60}{hgflightblue!60!white}
\colorlet{hgflightblue70}{hgflightblue!70!white}
\colorlet{hgflightblue80}{hgflightblue!80!white}
\colorlet{hgflightblue90}{hgflightblue!90!white}

\definecolor{hgfgreen}{RGB}{140, 180, 35}
\colorlet{hgfgreen10}{hgfgreen!10!white}
\colorlet{hgfgreen20}{hgfgreen!20!white}
\colorlet{hgfgreen30}{hgfgreen!30!white}
\colorlet{hgfgreen40}{hgfgreen!40!white}
\colorlet{hgfgreen50}{hgfgreen!50!white}
\colorlet{hgfgreen60}{hgfgreen!60!white}
\colorlet{hgfgreen70}{hgfgreen!70!white}
\colorlet{hgfgreen80}{hgfgreen!80!white}
\colorlet{hgfgreen90}{hgfgreen!90!white}

\definecolor{hgfgray}{RGB}{90, 105, 110}
\colorlet{hgfgray10}{hgfgray!10!white}
\colorlet{hgfgray20}{hgfgray!20!white}
\colorlet{hgfgray30}{hgfgray!30!white}
\colorlet{hgfgray40}{hgfgray!40!white}
\colorlet{hgfgray50}{hgfgray!50!white}
\colorlet{hgfgray60}{hgfgray!60!white}
\colorlet{hgfgray70}{hgfgray!70!white}
\colorlet{hgfgray80}{hgfgray!80!white}
\colorlet{hgfgray90}{hgfgray!90!white}

\definecolor{hgfhighlight}{RGB}{205, 238, 251}
\colorlet{hgfhighlight10}{hgfhighlight!10!white}
\colorlet{hgfhighlight20}{hgfhighlight!20!white}
\colorlet{hgfhighlight30}{hgfhighlight!30!white}
\colorlet{hgfhighlight40}{hgfhighlight!40!white}
\colorlet{hgfhighlight50}{hgfhighlight!50!white}
\colorlet{hgfhighlight60}{hgfhighlight!60!white}
\colorlet{hgfhighlight70}{hgfhighlight!70!white}
\colorlet{hgfhighlight80}{hgfhighlight!80!white}
\colorlet{hgfhighlight90}{hgfhighlight!90!white}

\definecolor{hgfmint}{RGB}{5, 229, 186}
\colorlet{hgfmint10}{hgfmint!10!white}
\colorlet{hgfmint20}{hgfmint!20!white}
\colorlet{hgfmint30}{hgfmint!30!white}
\colorlet{hgfmint40}{hgfmint!40!white}
\colorlet{hgfmint50}{hgfmint!50!white}
\colorlet{hgfmint60}{hgfmint!60!white}
\colorlet{hgfmint70}{hgfmint!70!white}
\colorlet{hgfmint80}{hgfmint!80!white}
\colorlet{hgfmint90}{hgfmint!90!white}

\definecolor{hgfpale}{RGB}{236, 251, 253}
\colorlet{hgfpale10}{hgfpale!10!white}
\colorlet{hgfpale20}{hgfpale!20!white}
\colorlet{hgfpale30}{hgfpale!30!white}
\colorlet{hgfpale40}{hgfpale!40!white}
\colorlet{hgfpale50}{hgfpale!50!white}
\colorlet{hgfpale60}{hgfpale!60!white}
\colorlet{hgfpale70}{hgfpale!70!white}
\colorlet{hgfpale80}{hgfpale!80!white}
\colorlet{hgfpale90}{hgfpale!90!white}

\definecolor{hgfaerospace}{RGB}{80, 200, 170}
\definecolor{hgfast}{named}{hgfaerospace}

\colorlet{hgfaerospace10}{hgfaerospace!10!white}
\colorlet{hgfaerospace20}{hgfaerospace!20!white}
\colorlet{hgfaerospace30}{hgfaerospace!30!white}
\colorlet{hgfaerospace40}{hgfaerospace!40!white}
\colorlet{hgfaerospace50}{hgfaerospace!50!white}
\colorlet{hgfaerospace60}{hgfaerospace!60!white}
\colorlet{hgfaerospace70}{hgfaerospace!70!white}
\colorlet{hgfaerospace80}{hgfaerospace!80!white}
\colorlet{hgfaerospace90}{hgfaerospace!90!white}

\colorlet{hgfast10}{hgfast!10!white}
\colorlet{hgfast20}{hgfast!20!white}
\colorlet{hgfast30}{hgfast!30!white}
\colorlet{hgfast40}{hgfast!40!white}
\colorlet{hgfast50}{hgfast!50!white}
\colorlet{hgfast60}{hgfast!60!white}
\colorlet{hgfast70}{hgfast!70!white}
\colorlet{hgfast80}{hgfast!80!white}
\colorlet{hgfast90}{hgfast!90!white}

\definecolor{hgfearthandenvironment}{RGB}{50, 100, 105}
\definecolor{hgfee}{named}{hgfearthandenvironment}

\colorlet{hgfearthandenvironment10}{hgfearthandenvironment!10!white}
\colorlet{hgfearthandenvironment20}{hgfearthandenvironment!20!white}
\colorlet{hgfearthandenvironment30}{hgfearthandenvironment!30!white}
\colorlet{hgfearthandenvironment40}{hgfearthandenvironment!40!white}
\colorlet{hgfearthandenvironment50}{hgfearthandenvironment!50!white}
\colorlet{hgfearthandenvironment60}{hgfearthandenvironment!60!white}
\colorlet{hgfearthandenvironment70}{hgfearthandenvironment!70!white}
\colorlet{hgfearthandenvironment80}{hgfearthandenvironment!80!white}
\colorlet{hgfearthandenvironment90}{hgfearthandenvironment!90!white}

\colorlet{hgfee10}{hgfee!10!white}
\colorlet{hgfee20}{hgfee!20!white}
\colorlet{hgfee30}{hgfee!30!white}
\colorlet{hgfee40}{hgfee!40!white}
\colorlet{hgfee50}{hgfee!50!white}
\colorlet{hgfee60}{hgfee!60!white}
\colorlet{hgfee70}{hgfee!70!white}
\colorlet{hgfee80}{hgfee!80!white}
\colorlet{hgfee90}{hgfee!90!white}

\definecolor{hgfenergy}{RGB}{255, 210, 40}

\colorlet{hgfenergy10}{hgfenergy!10!white}
\colorlet{hgfenergy20}{hgfenergy!20!white}
\colorlet{hgfenergy30}{hgfenergy!30!white}
\colorlet{hgfenergy40}{hgfenergy!40!white}
\colorlet{hgfenergy50}{hgfenergy!50!white}
\colorlet{hgfenergy60}{hgfenergy!60!white}
\colorlet{hgfenergy70}{hgfenergy!70!white}
\colorlet{hgfenergy80}{hgfenergy!80!white}
\colorlet{hgfenergy90}{hgfenergy!90!white}

\definecolor{hgfhealth}{RGB}{210, 50, 100}

\colorlet{hgfhealth10}{hgfhealth!10!white}
\colorlet{hgfhealth20}{hgfhealth!20!white}
\colorlet{hgfhealth30}{hgfhealth!30!white}
\colorlet{hgfhealth40}{hgfhealth!40!white}
\colorlet{hgfhealth50}{hgfhealth!50!white}
\colorlet{hgfhealth60}{hgfhealth!60!white}
\colorlet{hgfhealth70}{hgfhealth!70!white}
\colorlet{hgfhealth80}{hgfhealth!80!white}
\colorlet{hgfhealth90}{hgfhealth!90!white}

\definecolor{hgfinformation}{RGB}{160, 35, 90}
\definecolor{hginfo}{named}{hgfinformation}

\colorlet{hgfinformation10}{hgfinformation!10!white}
\colorlet{hgfinformation20}{hgfinformation!20!white}
\colorlet{hgfinformation30}{hgfinformation!30!white}
\colorlet{hgfinformation40}{hgfinformation!40!white}
\colorlet{hgfinformation50}{hgfinformation!50!white}
\colorlet{hgfinformation60}{hgfinformation!60!white}
\colorlet{hgfinformation70}{hgfinformation!70!white}
\colorlet{hgfinformation80}{hgfinformation!80!white}
\colorlet{hgfinformation90}{hgfinformation!90!white}

\colorlet{hgfinfo10}{hgfinformation!10!white}
\colorlet{hgfinfo20}{hgfinformation!20!white}
\colorlet{hgfinfo30}{hgfinformation!30!white}
\colorlet{hgfinfo40}{hgfinformation!40!white}
\colorlet{hgfinfo50}{hgfinformation!50!white}
\colorlet{hgfinfo60}{hgfinformation!60!white}
\colorlet{hgfinfo70}{hgfinformation!70!white}
\colorlet{hgfinfo80}{hgfinformation!80!white}
\colorlet{hgfinfo90}{hgfinformation!90!white}

\definecolor{hgfmatter}{RGB}{240, 120, 30}

\colorlet{hgfmatter10}{hgfmatter!10!white}
\colorlet{hgfmatter20}{hgfmatter!20!white}
\colorlet{hgfmatter30}{hgfmatter!30!white}
\colorlet{hgfmatter40}{hgfmatter!40!white}
\colorlet{hgfmatter50}{hgfmatter!50!white}
\colorlet{hgfmatter60}{hgfmatter!60!white}
\colorlet{hgfmatter70}{hgfmatter!70!white}
\colorlet{hgfmatter80}{hgfmatter!80!white}
\colorlet{hgfmatter90}{hgfmatter!90!white}

%% file: figures/graphical-abstract.tikz
\newcommand{\distplotcol}[4]{
    \begin{tikzpicture}
        \begin{axis}[
            xmin=-3, xmax=4.2,
            ymin=-0.1, ymax=1.0,
            xtick=\empty,
            ytick=\empty,
            xticklabel=\empty,
            yticklabel=\empty,
            axis x line=bottom,
            x axis line style={
                thick,
                draw=black,
                {-Stealth[width=1mm, length=1mm]}
            },
            axis y line=left,
            y axis line style={
                thick,
                draw=black,
                {-Stealth[width=1mm, length=1mm]}
            },
            clip=true,
            width=#4,
            height=#4,
            enlarge x limits=false,
]
            \addplot[no marks, smooth, #3, samples=100, domain=-3.:3.0] {(2*pi*#2^2)^(-1/2)*e^(-(x-#1)^2/(2*#2^2))};
        \end{axis}
    \end{tikzpicture}
}

\newcommand{\makeshadows}[3]{
    \begin{scope}[canvas is xz plane at y=0.0]
        \node[anchor=north, inner sep=0, draw=#1, fill=#2, minimum width=3cm, minimum height=0.2cm, inner sep=0, transform shape] at (1.5, -4.0) {};

        \node[anchor=north, inner sep=0, draw=#1, fill=#3, minimum width=1.0cm, minimum height=0.1cm, transform shape] at (-1.15, -2.0) {};

        \node[anchor=north, inner sep=0, draw=#1, fill=#3, minimum width=1.0cm, minimum height=0.1cm, transform shape] at (0.15, -2.0) {};
    
        \node[anchor=north, inner sep=0, draw=#1, fill=#3, minimum width=1.0cm, minimum height=0.1cm, transform shape] at (-0.05, -3.1) {};

        \node[anchor=north, inner sep=0, draw=#1, fill=#3, minimum width=1.0cm, minimum height=0.1cm, transform shape] at (1.25, -3.1) {};
    \end{scope}

    \begin{scope}[canvas is yz plane at x=0.0]
        \node[anchor=north, inner sep=0, draw=#1, fill=#2, minimum width=3cm, minimum height=0.2cm, inner sep=0, transform shape] at (1.5, -4.0) {};
    
        \node[anchor=north, inner sep=0, draw=#1, fill=#3, minimum width=0.6cm, minimum height=0.1cm, transform shape] at (1.95, -3.65) {};
        
        \node[anchor=north, inner sep=0, draw=#1, fill=#3, minimum width=0.6cm, minimum height=0.1cm, transform shape] at (0.65, -2.35) {};
        
        \node[anchor=north, inner sep=0, draw=#1, fill=#3, minimum width=0.6cm, minimum height=0.1cm, transform shape] at (0.85, -3.65) {};
        
        \node[anchor=north, inner sep=0, draw=#1, fill=#3, minimum width=0.6cm, minimum height=0.1cm, transform shape] at (-0.45, -2.35) {};
    \end{scope}
}


\begin{tikzpicture}[
    isometric view,
    font=\sffamily\scriptsize
]
    \begin{scope}
        \node[fill=white, minimum width=0.33\linewidth, minimum height=7cm] (dtp-box) {};
        \node[anchor=north, below=0.2 of dtp-box.north] {\bfseries Domain--Tensor-Parallelism (DTP)};

        \begin{scope}[canvas is xy plane at z=-2.5]
            \node[draw=hgfmint, fill=hgfmint10, minimum width=3cm, minimum height=3cm, inner sep=0, text height=2.6cm, transform shape] (node) {};
            \node[text height=2.6cm, transform shape] at ($(node) + (0.0, 0.1)$) {\bfseries \bfseries DTP instance};
    
            \node[draw=hgfmint, fill=hgfmint30, minimum width=1cm, minimum height=0.6cm, xshift=-0.65cm, yshift=0.8cm, transform shape] (gpu1) {\bfseries GPU 1};
            \node[draw=hgfmint, fill=hgfmint30, minimum width=1cm, minimum height=0.6cm, xshift=0.65cm, yshift=0.8cm, transform shape] (gpu2) {\bfseries GPU 2};
            \node[draw=hgfmint, fill=hgfmint30, minimum width=1cm, minimum height=0.6cm, xshift=-0.65cm, yshift=-0.3cm, transform shape] (gpu3) {\bfseries GPU 3};
            \node[draw=hgfmint, fill=hgfmint30, minimum width=1cm, minimum height=0.6cm, xshift=0.65cm, yshift=-0.3cm, transform shape] (gpu4) {\bfseries GPU 4};
        \end{scope}

        
        \makeshadows{hgfmint}{hgfmint30}{hgfmint50}

        \begin{scope}[canvas is xz plane at y=3]
            \node[draw=black, inner sep=0, thick, transform shape] at (0.2, -2.5) {\includegraphics[scale=0.8]{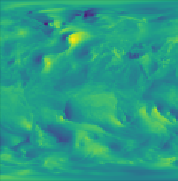}};
            \node[draw=black, inner sep=0, thick, transform shape] (bottom-back-left) at (0.3, -2.6) {\includegraphics[scale=0.8]{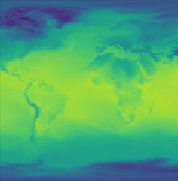}};
            
            \node[draw=black, inner sep=0, thick, transform shape] at (1.8, -2.5) {\includegraphics[scale=0.8]{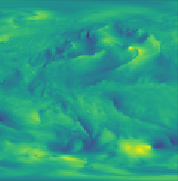}};
            \node[draw=black, inner sep=0, thick, transform shape]  at (1.9, -2.6) {\includegraphics[scale=0.8]{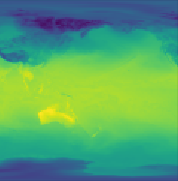}};
            
            \node[draw=black, inner sep=0, thick, transform shape] (top-back-left) at (0.2, -0.9) {\includegraphics[scale=0.8]{images/v10m-left.png}};
            \node[draw=black, inner sep=0, thick, transform shape] at (0.3, -1.0) {\includegraphics[scale=0.8]{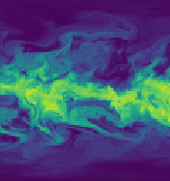}};
            
            \node[draw=black, inner sep=0, thick, transform shape] (top-back-right) at (1.8, -0.9) {\includegraphics[scale=0.8]{images/v10m-right.png}};
            \node[draw=black, inner sep=0, thick, transform shape] at (1.9, -1.0) {\includegraphics[scale=0.8]{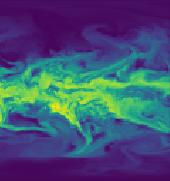}};

            \draw[stealth-stealth, shorten <= 0.1cm, shorten >= 0.1cm] (top-back-left.north) |-|[distance=0.4cm] (top-back-right.north) node[midway, above, transform shape] {Spatial parallel};
            
            \draw[stealth-stealth, shorten <= 0.1cm, shorten >= 0.12cm] (top-back-left.west) -|-[distance=-0.4cm] (bottom-back-left.west) node[midway, above, transform shape, rotate=90] {Channel parallel};
        \end{scope}

        \node[inner sep=0] at (-0.25, 0, 1.2) {\bfseries\faTimes};
    \end{scope}

    \begin{scope}[xshift=0.33\linewidth]
        \node[fill=white, minimum width=0.31\linewidth, minimum height=7cm] (up-box) {};
        \node[anchor=north, below=0.2 of up-box.north] {\bfseries Uncertainty-Parallelism (UP)};

        \begin{scope}[canvas is xy plane at z=-2.5]
            \node[draw=hgfblue, fill=hgfblue10, minimum width=3cm, minimum height=3cm, inner sep=0, text height=2.6cm, transform shape] (up-group) {};
            \node[text height=2.6cm, transform shape] at ($(up-group) + (0.0, 0.1)$) {\bfseries \bfseries UP instance};
    
            \node[inner sep=0, draw=hgfblue, fill=hgfblue20, minimum width=1cm, minimum height=0.6cm, xshift=-0.65cm, yshift=0.8cm, transform shape] (node1) {\bfseries DTP 1};
            \node[inner sep=0, draw=hgfblue, fill=hgfblue20, minimum width=1cm, minimum height=0.6cm, xshift=0.65cm, yshift=0.8cm, transform shape] (node2) {\bfseries DTP 2};
            \node[inner sep=0, draw=hgfblue, fill=hgfblue20, minimum width=1cm, minimum height=0.6cm, xshift=-0.65cm, yshift=-0.3cm, transform shape] (node3) {\bfseries DTP 3};
            \node[inner sep=0, draw=hgfblue, fill=hgfblue20, minimum width=1cm, minimum height=0.6cm, xshift=0.65cm, yshift=-0.3cm, transform shape] (node4) {\bfseries DTP 4};
        \end{scope}


        \makeshadows{hgfblue}{hgfblue30}{hgfblue40}

        \begin{scope}[canvas is xz plane at y=3]
            \node[draw=none, inner sep=0, transform shape, anchor=center] (aleatoric-label) at (1.0, 0.0) {Aleatoric / noise injection};
        
            \node[draw=black, inner sep=0, transform shape] at (0.3, -2.6) {\includegraphics[width=0.07\linewidth]{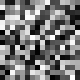}};
            
            \node[draw=black, inner sep=0, transform shape] at (1.9, -2.6) {\includegraphics[width=0.07\linewidth]{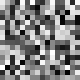}};
            
            \node[draw=black, inner sep=0, transform shape] at (0.3, -1.0) {\includegraphics[width=0.07\linewidth]{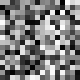}};
            
            \node[draw=black, inner sep=0, transform shape] at (1.9, -1.0) {\includegraphics[width=0.07\linewidth]{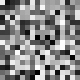}};
        \end{scope}
    \end{scope}


    \begin{scope}[xshift=0.67\linewidth]
        \node[fill=white, minimum width=0.33\linewidth, minimum height=7cm] (ddp-box) {};
        \node[anchor=north, below=0.2 of ddp-box.north] {\bfseries Data-Parallelism (DP)};
        
        \begin{scope}[canvas is xy plane at z=-2.5]
            \node[draw=hgfdarkblue, fill=hgfdarkblue20, minimum width=3cm, minimum height=3cm, inner sep=0, text height=2.6cm, transform shape] (cluster) {};
            \node[text height=2.6cm, transform shape] at ($(cluster) + (0.0, 0.1)$) {\bfseries \bfseries DP instance};
    
            \node[draw=hgfblue, fill=hgfdarkblue30, minimum width=1cm, minimum height=0.6cm, xshift=-0.65cm, yshift=0.8cm, transform shape] (up1) {\bfseries UP 1};
            \node[draw=hgfblue, fill=hgfdarkblue30, minimum width=1cm, minimum height=0.6cm, xshift=0.65cm, yshift=0.8cm, transform shape] (up2) {\bfseries UP 2};
            \node[draw=hgfblue, fill=hgfdarkblue30, minimum width=1cm, minimum height=0.6cm, xshift=-0.65cm, yshift=-0.3cm, transform shape] (up3) {\bfseries UP 3};
            \node[draw=hgfblue, fill=hgfdarkblue30, minimum width=1cm, minimum height=0.6cm, xshift=0.65cm, yshift=-0.3cm, transform shape] (up4) {\bfseries UP 4};
        \end{scope}


        \makeshadows{hgfdarkblue}{hgfdarkblue40}{hgfdarkblue50}

        
        \foreach \z [count=\i] in {0.5, 0.7, ..., 1.5} {
            \node[draw=black, inner sep=0, canvas is xy plane at z=\z, transform shape] (data-bottom-right-\i) at (-1.5, -0.5) {\includegraphics[scale=0.4]{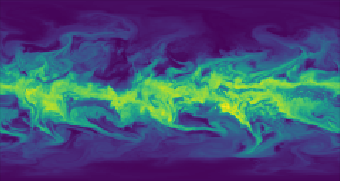}};
        }
        \foreach \z [count=\i] in {2.5, 2.7, ..., 3.5} {
            \node[draw=black, inner sep=0, canvas is xy plane at z=\z, transform shape] (data-top-right-\i) at (-1.5, -0.5) {\includegraphics[scale=0.4]{images/q850.png}};
        }
        \foreach \z [count=\i] in {0.5, 0.7, ..., 1.5} {
            \node[draw=black, inner sep=0, canvas is xy plane at z=\z, transform shape] (data-bottom-left-\i) at (-3.0, -0.5) {\includegraphics[scale=0.4]{images/q850.png}};
        }
        \foreach \z [count=\i] in {2.5, 2.7, ..., 3.5} {
            \node[draw=black, inner sep=0, canvas is xy plane at z=\z, transform shape] (data-top-left-\i) at (-3.0, -0.5) {\includegraphics[scale=0.4]{images/q850.png}};
        }
        
        \begin{scope}[canvas is xz plane at y=3]
            \draw[stealth-stealth, shorten <= 0.1cm, shorten >= 0.1cm] (data-top-left-4.north west) -|-[distance=-0.4cm] (data-bottom-left-3.north west) node[midway, above, transform shape, rotate=90] {Data parallel};
        \end{scope}

        \node[align=left, text width=0.15\linewidth, anchor=west] at (0.8, 0.5) {
            \textbf{ERA5 Hourly Data}\\
            Time frame: 1979--2019\\
            Format: zarr, chunked in time and space
        };
    \end{scope}

    \draw[hgfgray, densely dotted] (node1.north east) -- (node.north east);
    \draw[hgfgray, densely dotted] (node1.south west) -- (node.south west);
    \draw[hgfgray, densely dotted] (up1.north east) -- (up-group.north east);
    \draw[hgfgray, densely dotted] (up1.south west) -- (up-group.south west);

    \begin{scope}[canvas is yz plane at x=3]
        \node[fill=white, xscale=-1, transform shape] (weight-matrix) at (1.2, -1.8) {$\begin{pmatrix}
            \mathbf{W}_{1,1} & \cdots & \cdots & \mathbf{W}_{1,m} \\
            \vdots & \ddots & & \vdots\\
            & & & \\
            \vdots & & \ddots & \vdots \\
            \mathbf{W}_{n,1} & \cdots & \cdots & \mathbf{W}_{n,m}
        \end{pmatrix}$};

        \draw[hgfgray, thick, densely dotted] (weight-matrix.north) -- (weight-matrix.south);
        
        \draw[stealth-stealth, shorten <= 0.1cm, shorten >= 0.1cm] ([xshift=-0.3cm]weight-matrix.north west) |-|[distance=-0.5cm] ([xshift=0.3cm]weight-matrix.north east) node[midway, above, xscale=-1, transform shape] {Tensor parallel};
    \end{scope}
    
    \node[inner sep=0] at (-0.25, 0, 1.2) {\bfseries\faTimes};

    \begin{scope}[xshift=0.33\linewidth]
        \begin{scope}[canvas is yz plane at x=3, xscale=-1, xshift=-2.0cm]
            \node[draw=none, fill=white, inner sep=0, transform shape, anchor=center] (epistemic-label) at (1.0, 0.0) {Epistemic / weight sampling};
        
            \node[draw=none, fill=white, inner sep=0, transform shape] at (0.3, -2.6) {\distplotcol{0.5}{0.5}{hgfblue50}{0.16\linewidth}};
            
            \node[draw=none, fill=white, inner sep=0, transform shape] at (1.9, -2.6) {\distplotcol{0.0}{1.3}{hgfblue50}{0.16\linewidth}};
            
            \node[draw=none, fill=white, inner sep=0, transform shape] at (0.3, -1.0) {\distplotcol{-1.0}{0.6}{hgfblue50}{0.16\linewidth}};
            
            \node[draw=none, fill=white, inner sep=0, transform shape] at (1.9, -1.0) {\distplotcol{0.0}{0.5}{hgfblue50}{0.16\linewidth}};
        \end{scope}
    \end{scope}
    
    \begin{scope}[canvas is yz plane at x=0.0]

    \end{scope}
\end{tikzpicture}

%% file: figures/architecture.tikz
\begin{tikzpicture}[
    isometric view,
    font=\sffamily\scriptsize
]
    \node[draw=black, inner sep=0] (input-t-back) {\includegraphics[scale=0.6]{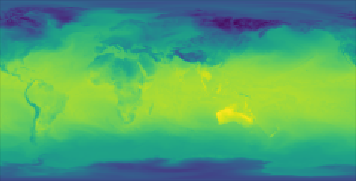}};
    \node[draw=black, inner sep=0, xshift=0.1cm, yshift=-0.1cm] (input-t-mid) at (input-t-back) {\includegraphics[scale=0.6]{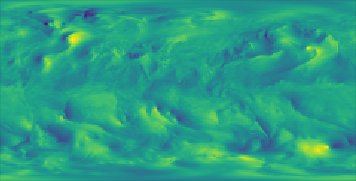}};
    \node[draw=black, inner sep=0, xshift=0.1cm, yshift=-0.1cm] (input-t-front) at (input-t-mid) {\includegraphics[scale=0.6]{images/q850.png}};
    \node[below=0.1 of input-t-mid] {Input at time $t$};
    
    \node[draw=black, inner sep=0, below=1.3 of input-t-back] (input-t-1-back) {\includegraphics[scale=0.6]{images/t2m.png}};
    \node[draw=black, inner sep=0, xshift=0.1cm, yshift=-0.1cm] (input-t-1-mid) at (input-t-1-back) {\includegraphics[scale=0.6]{images/v10m.png}};
    \node[draw=black, inner sep=0, xshift=0.1cm, yshift=-0.1cm] (input-t-1-front) at (input-t-1-mid) {\includegraphics[scale=0.6]{images/q850.png}};
    \node[below=0.1 of input-t-1-mid] {Input at time $t - \Delta t$};

    \draw[-] (input-t-front.east) -|-[distance=0.5cm] (input-t-1-front.east) node[midway, circle, draw=black, fill=white, inner sep=0] (concatenate) {\bfseries\small\faPlus};
    \node[below=0.1 of concatenate, fill=white] {concat.};

    \node[draw=black, fill=hgfblue20, minimum height=1.0cm, minimum width=1.7cm, align=center, right=1.5cm of input-t-front] (patch-embedding) {\bfseries Patch Embed.\\2$\times$2 Conv2D};
    \draw[-stealth] (concatenate) -|- (patch-embedding);

    \node[draw=black, fill=hgfblue20, minimum height=1.0cm, minimum width=1.7cm, align=center, right=0.5cm of patch-embedding] (swin-1) {\bfseries Swin\\\bfseries Transformer};
    \draw[-stealth] (patch-embedding) -- (swin-1) node[midway, inner sep=0] (residual-start) {};

    \node[draw=black, fill=hgfblue20, minimum height=1.0cm, minimum width=1.7cm, align=center, right=0.5cm of swin-1] (downsample) {\bfseries Downsample\\2$\times$2 Conv2D};
    \draw[-stealth] (swin-1) -- (downsample);

    \node[draw=black, fill=hgfblue20, minimum height=1.0cm, minimum width=1.7cm, align=center, right=0.5cm of downsample] (swin-2) {\bfseries Swin\\\bfseries Transformer};
    \draw[-stealth] (downsample) -- (swin-2);

    \node[draw=black, fill=hgfblue20, minimum height=1.0cm, minimum width=1.7cm, align=center, right=0.5cm of swin-2] (upsample) {\bfseries Upsample\\2$\times$2 Conv2DT};
    \draw[-stealth] (swin-2) -- (upsample);

    \node[draw=black, fill=hgfblue20, minimum height=1.0cm, minimum width=1.7cm, align=center, right=0.5cm of upsample] (swin-3) {\bfseries Swin\\\bfseries Transformer};
    \draw[-stealth] (upsample) -- (swin-3);

    \node[draw=black, fill=hgfblue20, minimum height=1.0cm, minimum width=1.7cm, align=center, below=1.0cm of swin-3] (patch-recovery) {\bfseries Patch Recov.\\2$\times$2 Conv2DT};
    \draw[-stealth] (swin-3) -- (patch-recovery) node[circle, draw=black, fill=white, midway, inner sep=0] (residual-end) {\bfseries\small\faPlus};
    \draw[-stealth] (residual-start) |- (residual-end) node[pos=0.75, above] {concat. along variables};
    
    \node[draw=black, inner sep=0, above left=0.2cm and 0.7cm of patch-recovery.west, anchor=east] (output-t-1-back) {\includegraphics[scale=0.6]{images/t2m.png}};
    \node[draw=black, inner sep=0, xshift=0.1cm, yshift=-0.1cm] (output-t-1-mid) at (output-t-1-back) {\includegraphics[scale=0.6]{images/v10m.png}};
    \node[draw=black, inner sep=0, xshift=0.1cm, yshift=-0.1cm] (output-t-1-front) at (output-t-1-mid) {\includegraphics[scale=0.6]{images/q850.png}};
    \node[below=0.1 of output-t-1-mid] {Output at time $t+\Delta t$};
    \draw[-stealth] (patch-recovery) -- (output-t-1-front);

    \node[draw=black, inner sep=0, above left=0.2cm and 4cm of patch-recovery.west, anchor=east] (output-t-2-back) {\includegraphics[scale=0.6]{images/t2m.png}};
    \node[draw=black, inner sep=0, xshift=0.1cm, yshift=-0.1cm] (output-t-2-mid) at (output-t-2-back) {\includegraphics[scale=0.6]{images/v10m.png}};
    \node[draw=black, inner sep=0, xshift=0.1cm, yshift=-0.1cm] (output-t-2-front) at (output-t-2-mid) {\includegraphics[scale=0.6]{images/q850.png}};
    \node[below=0.1 of output-t-2-mid] {Output at time $t+2\Delta t$};
    \draw (output-t-2-front.east) -- ($(output-t-1-front.west)-(0.235cm, 0)$) node[circle, draw=black, fill=white, midway, inner sep=0] {\bfseries\small\faPlus};

    \node[fill=hgfgray!2!white, draw=black, below=1.1 of input-t-1-back.south west, anchor=north west, minimum width=0.93\linewidth, minimum height=5.6cm] (dtp-back) {};
    \node[fill=hgfgray!2!white, draw=black, xshift=0.2cm, yshift=-0.2cm, minimum width=0.93\linewidth, minimum height=5.6cm] (dtp) at (dtp-back) {};
    \node[above=0.1 of dtp.south, anchor=south] {\bfseries Swin Transformer Block \mdseries (windows in parallel)};
    \draw[hgfgray, densely dotted] ([yshift=0.1cm]dtp.west) -- ([yshift=0.1cm]dtp.east);

    \node[draw=black, fill=hgfblue20, minimum height=1.7cm, minimum width=1.0cm, align=center, below right=0.8cm and 1.4cm of dtp.north west, anchor=north west] (input-transformer-1) {\bfseries Local\\\bfseries input};
    \node[draw=black, fill=hgfblue20, minimum height=1.7cm, minimum width=1.0cm, align=center, below=0.5 of input-transformer-1] (input-transformer-2) {\bfseries Local\\\bfseries input};
    \draw[|-stealth] ([yshift=0.2cm]input-transformer-1.north west) -- ([yshift=0.2cm]input-transformer-1.north east) node[midway, above] {local channels};
    \draw[|-stealth] ([xshift=-0.2cm]input-transformer-1.north west) -- ([xshift=-0.2cm]input-transformer-1.south west) node[midway, above, rotate=90] {seq.};

    \node[left=0.8cm of input-transformer-1.west, , anchor=south, rotate=90] {\bfseries GPU 1};
    \node[left=0.8cm of input-transformer-2.west, , anchor=south, rotate=90] {\bfseries GPU 2};

    \node[draw=black, fill=hgfmint50, minimum width=1.0cm, minimum height=3.82cm, align=center, right=0.6cm of input-transformer-1.north east, anchor=north west] (dist-mlp-1) {\bfseries Dist.\\\bfseries Linear\\~\\P2P\\ comm.};

    \draw[-stealth] (input-transformer-1.east) -- (input-transformer-1.east -| dist-mlp-1.west);
    \draw[-stealth] (input-transformer-2.east) -- (input-transformer-2.east -| dist-mlp-1.west);
    
    \node[draw=black, fill=hgfblue20, minimum height=1.3cm, minimum width=0.9cm, right=0.8cm of dist-mlp-1.north east, anchor=north west] (q) {};
    \node[anchor=south west, yshift=-0.2cm] at (q.north east) {\bfseries Q};
    \node[draw=black, fill=hgfblue20, minimum height=1.3cm, minimum width=0.9cm, xshift=0.2cm, yshift=-0.2cm] (k) at (q) {};
    \node[anchor=south west, yshift=-0.2cm] at (k.north east) {\bfseries K};
    \node[draw=black, fill=hgfblue20, minimum height=1.3cm, minimum width=0.9cm, xshift=0.2cm, yshift=-0.2cm] (v) at (k) {};
    \node[anchor=south west, yshift=-0.2cm] at (v.north east) {\bfseries V};
    
    \draw[|-stealth] ([yshift=0.2cm]q.north west) -- ([yshift=0.2cm]q.north east) node[midway, above] {Embed. dim / c};
    \draw[|-stealth] ([xshift=-0.2cm]q.north west) -- ([xshift=-0.2cm]q.south west) node[midway, above, rotate=90] {\hspace{1.5em}seq.};

    \draw[-stealth, shorten >= 0.5cm] (dist-mlp-1.east |- k.west) -- (k.west);

    \node[draw=black, fill=hgfblue20, minimum height=1.3cm, minimum width=0.9cm, above right=0.4cm and 0.8cm of dist-mlp-1.south east, anchor=south west] (q-2) {};
    \node[anchor=south west, yshift=-0.2cm] at (q-2.north east) {\bfseries Q};
    \node[draw=black, fill=hgfblue20, minimum height=1.3cm, minimum width=0.9cm, xshift=0.2cm, yshift=-0.2cm] (k-2) at (q-2) {};
    \node[anchor=south west, yshift=-0.2cm] at (k-2.north east) {\bfseries K};
    \node[draw=black, fill=hgfblue20, minimum height=1.3cm, minimum width=0.9cm, xshift=0.2cm, yshift=-0.2cm] (v-2) at (k-2) {};
    \node[anchor=south west, yshift=-0.2cm] at (v-2.north east) {\bfseries V};

    \draw[-stealth, shorten >= 0.2cm] (dist-mlp-1.east |- k-2.west) -- (k-2.west);

    \node[draw=black, fill=hgfblue10, above right=0.3cm and 1.0cm of q.north east, anchor=north west, minimum height=1.8cm, minimum width=5.0cm] (flash-attention-back) {};
    \node[draw=black, fill=hgfblue10, xshift=0.2cm, yshift=-0.2cm, minimum height=1.8cm, minimum width=5.0cm] (flash-attention-front) at (flash-attention-back) {
        softmax$\begin{pmatrix}
        \qquad & \qquad & \qquad \\
        \qquad & \qquad & \qquad \\
        \qquad & \qquad & \qquad \\
        \qquad & \qquad & \qquad \\
        \qquad & \qquad & \qquad \\
        \end{pmatrix}\times$\hspace{0.8cm}
    };
    \draw[-stealth] (v.east) -- (v.east -| flash-attention-front.west);
    
    \node[draw=black, fill=hgfblue20, minimum height=1.2cm, minimum width=0.5cm] at ([xshift=-0.8cm]flash-attention-front.center) {Q};
    \node[draw=black, fill=hgfblue20, minimum height=0.5cm, minimum width=1.2cm] at ([xshift=0.3cm]flash-attention-front.center) {K\textsuperscript{T}};
    \node[draw=black, fill=hgfblue20, minimum height=1.2cm, minimum width=0.5cm] at ([xshift=1.9cm]flash-attention-front.center) {V};
    
    \draw[|-stealth] ([xshift=0.2cm]flash-attention-back.north east) -- ([xshift=0.2cm]flash-attention-front.north east) node[midway, xshift=0.5cm, yshift=0.2cm] {$\times$ heads};

    \node[above=0.2 of flash-attention-front, anchor=south] {\bfseries FlashAttention-2 / Transformer Engine};

    \node[draw=black, fill=hgfblue10, right=1.0cm of q-2.north east, anchor=north west, minimum height=1.8cm, minimum width=5.0cm] (flash-attention-back-2) {};
    \node[draw=black, fill=hgfblue10, xshift=0.2cm, yshift=-0.2cm, minimum height=1.8cm, minimum width=5.0cm] (flash-attention-front-2) at (flash-attention-back-2) {
        softmax$\begin{pmatrix}
        \qquad & \qquad & \qquad \\
        \qquad & \qquad & \qquad \\
        \qquad & \qquad & \qquad \\
        \qquad & \qquad & \qquad \\
        \qquad & \qquad & \qquad \\
        \end{pmatrix}\times$\hspace{0.8cm}
    };
    \draw[-stealth] (v-2.east) -- (v-2.east -| flash-attention-front-2.west);
    
    \node[draw=black, fill=hgfblue20, minimum height=1.2cm, minimum width=0.5cm] at ([xshift=-0.8cm]flash-attention-front-2.center) {Q};
    \node[draw=black, fill=hgfblue20, minimum height=0.5cm, minimum width=1.2cm] at ([xshift=0.3cm]flash-attention-front-2.center) {K\textsuperscript{T}};
    \node[draw=black, fill=hgfblue20, minimum height=1.2cm, minimum width=0.5cm] at ([xshift=1.9cm]flash-attention-front-2.center) {V};
    
    \node[draw=black, fill=hgfmint50, minimum width=1.0cm, minimum height=3.82cm, align=center, below right=0.1cm and 0.6cm of flash-attention-front.north east, anchor=north west] (dist-mlp-2) {\bfseries Dist.\\\bfseries Linear\\~\\P2P\\ comm.};

    \draw[-stealth] (flash-attention-front.east) -- (flash-attention-front.east -| dist-mlp-2.west);
    \draw[-stealth] (flash-attention-front-2.east) -- (flash-attention-front-2.east -| dist-mlp-2.west) node[midway, align=center] {};

    \node[draw=black, fill=hgfblue20, minimum height=1.7cm, minimum width=1.0cm, align=center, right=0.8cm of dist-mlp-2.north east, anchor=north west] (output-transformer-1) {\bfseries Local\\\bfseries output};
    \node[draw=black, fill=hgfblue20, minimum height=1.7cm, minimum width=1.0cm, align=center, below=0.5 of output-transformer-1] (output-transformer-2) {\bfseries Local\\\bfseries output};
    \draw[|-stealth] ([yshift=0.2cm]output-transformer-1.north west) -- ([yshift=0.2cm]output-transformer-1.north east) node[midway, above] {local channels};
    \draw[|-stealth] ([xshift=0.2cm]output-transformer-1.north east) -- ([xshift=0.2cm]output-transformer-1.south east) node[midway, below, rotate=90] {seq.};

    \draw[-stealth] (dist-mlp-2.east |- output-transformer-1.west) -- (output-transformer-1.west);
    \draw[-stealth] (dist-mlp-2.east |- output-transformer-2.west) -- (output-transformer-2.west);

    \draw[Circle-Circle, black, densely dotted, shorten >= -0.05cm, shorten <= -0.05cm] (swin-1.south) -- ([yshift=0.1cm]swin-1 |- residual-end) arc (90:-90:0.1cm) -- (swin-1.south |- dtp-back.north);
\end{tikzpicture}

%% file: plots/throughput-scaling.tikz
\begin{tikzpicture}[baseline={(0,0)}]
    \sffamily
    \begin{axis}[
        gbplot,
        width=\linewidth,
        height=0.6\linewidth,
        xlabel=GPUs,
        xmin=8,
        xmax=16384,
        xmode=log,
        xtick={16, 64, 256, 1024, 4096, 16384},
        xticklabels={16, 64, 256, 1024, 4096, 16\,384},
        log basis x=2,
        log ticks with fixed point,
        ylabel=Throughput (TFLOPS/s/GPU),
        legend style={
            legend columns=3,
            font=\scriptsize\bfseries,
        },
        ymin=1.0,
        ymax=380.0,
        ytick={50, 100, 150, 200, 250, 300, 350},
    ]
        \filldraw [fill=hgfgray20, draw=none] (axis cs:0.0, 0.0) rectangle (axis cs:192,600.0);

        \node[anchor=south east, align=right, font=\scriptsize] at (axis cs: 192, 300.0) {DP=1\\$\leftarrow$};
        \node[anchor=south west, align=left, font=\scriptsize] at (axis cs: 192, 300.0) {DP=16\\$\rightarrow$};
        
        \addplot[
            color=hgfblue,
            mark=square*,
            mark options={scale=0.7},
        ] table [
            col sep=comma,
            x=gpus,
            y=peak_tflops_gpu,
        ] {data/up_hero_table_jupiter.csv};

        \addplot[
            color=hgfinformation,
            mark=square*,
            mark options={scale=0.7},
        ] table [
            col sep=comma,
            x=gpus,
            x expr=\thisrow{gpus} / 2,
            y expr=\thisrow{peak_tflops_gpu} * 2,
            y=peak_tflops_gpu,
        ] {data/up_hero_table_lumi.csv};

        \node[anchor=south east, align=left, font=\scriptsize] at (axis cs: 16384, 320.0) {$\uparrow$ higher is better};
    \end{axis}
\end{tikzpicture}

%% file: plots/dp-scaling.tikz
\begin{tikzpicture}[baseline={(0,0)}]
    \sffamily
    \begin{axis}[
        gbplot,
        width=\linewidth,
        height=0.6\linewidth,
        xlabel=Nodes,
        xmin=2,
        xmax=6144,
        log ticks with fixed point,
        xmode=log,
        log basis x=2,
        xtick={4,16,64,256,1024,4096},
        xticklabels={4,16,64,256,1024,4096},
        ylabel=Peak performance (PFLOPS/s),
        ymode=log,
        log basis y=2,
        ymin=0.7,
        ymax=10000.0,
        ytick={1.0, 3.0, 10.0, 30.0, 100.0, 300.0, 1000.0, 3000.0, 10000.0},
        legend style={
            legend columns=3,
            font=\scriptsize\bfseries,
        },
        legend to name=leg:hero-legend
    ]
        \filldraw [fill=hgfgray20, draw=none] (axis cs:0.1, 0.1) rectangle (axis cs:48,12000.0);

        \node[anchor=south east, align=right, font=\scriptsize] at (axis cs: 48, 0.4) {UP=2\\$\leftarrow$};
        \node[anchor=south west, align=left, font=\scriptsize] at (axis cs: 48, 0.4) {UP=16\\$\rightarrow$};

        \node[anchor=east, align=left, font=\scriptsize\bfseries] at (axis cs: 5120, 5000) {5120 Nodes~$\rightarrow$};
        
        \addplot[ 
            color=hgfgray,
            densely dashed
        ] coordinates {
            (1, 1.0155390901418144)
            (8192, 8319.296226441744)
        };
        \addlegendentry{Ideal scaling};

        \addplot[ 
            forget plot,
            color=hgfgray,
            densely dashed
        ] coordinates {
            (1, 0.4912130493447488)
            (8192, 4024.017300232182)
        };
        
        \addplot[
            color=hgfblue,
            mark=square*,
            mark options={scale=0.7},
        ] table [
            col sep=comma,
            x=nodes,
            y expr={\thisrow{peak_tflops} / 1000.0}
        ] {data/dp_hero_table_jupiter.csv};
        \addlegendentry{\largejup};

        \addplot[
            color=hgfinformation,
            mark=square*,
            mark options={scale=0.7},
        ] table [
            col sep=comma,
            x=nodes,
            y expr={\thisrow{peak_tflops} / 1000.0}
        ] {data/dp_hero_table_lumi.csv};
        \addlegendentry{\largelumi};

        \node[anchor=south east, align=left, font=\scriptsize] at (axis cs: 6144, 2.5) {$\uparrow$ higher is better};
    \end{axis}
\end{tikzpicture}

%% file: plots/sample-scaling.tikz
\begin{tikzpicture}[baseline={(0,0)}]
    \sffamily
    \begin{axis}[
        gbplot,
        width=\linewidth,
        height=0.6\linewidth,
        log ticks with fixed point,
        xlabel=GPUs,
        xmin=7,
        xmax=2048,
        xmode=log,
        xtick={8, 32, 128, 512, 2048},
        xticklabels={32, 128, 512, 2\,048, 8\,192},
        ylabel=Samples (/s),
        ymode=log,
        log basis y=2,
        ymin=6,
        ymax=8000,
        ytick={4,16, 64, 256, 1024, 4096},
        yticklabels={4,16, 64, 256, 1024, 4096},
        legend style={
            legend columns=2,
            font=\scriptsize\bfseries,
        },
        legend to name=leg:sample-legend
    ]   

        
        \addplot[
            forget plot,
            color=hgfgray, 
            densely dashed,
        ] coordinates {
            (1.0, 1.192401825215598)
            (4096.0, 4884.077876083087)
        };

        \addplot[
            forget plot,
            color=hgfgray, 
            densely dashed,
        ] coordinates {
            (1.0, 3.403817535934437)
            (4096.0, 13942.03662718745)
        };
        
        \addplot[
            color=hgfmatter,
            mark=square*,
            mark options={scale=0.7},
        ] table [
            col sep=comma,
            x=nodes,
            y=samples_s,
            restrict x to domain={0:14}
        ] {data/predictive-training-table-jupiter.csv};
        \addlegendentry{\smalljup~ Training};
        
        \addplot[
            color=hgfee,
            mark=square*,
            mark options={scale=0.7},
        ] table [
            col sep=comma,
            x=nodes,
            y=samples_s,
            restrict x to domain={0:14}
        ] {data/predictive-inference-table-jupiter.csv};
        \addlegendentry{\smalljup~ Inference};
    \end{axis}
\end{tikzpicture}

%% file: plots/forecast-scaling.tikz
\begin{tikzpicture}[baseline={(0,0)}]
    \sffamily
    \begin{axis}[
        gbplot,
        width=\linewidth,
        height=0.6\linewidth,
        xlabel=DP Group Size,
        xmin=0.95,
        xmax=192,
        log ticks with fixed point,
        xmode=log,
        log basis x=2,
        xtick={1, 2, 4, 8, 16, 32, 64, 128, 256},
        xticklabels={1,2,4,8, 16,32,64,128, 256},
        ylabel=Forecasts (/s),
        ymode=log,
        log basis y=2,
        ymin=0.15,
        ymax=100,
        ytick={0.25, 0.5, 1, 2, 4, 8, 16, 32, 64, 128},
        legend style={
            legend columns=3,
            font=\scriptsize\bfseries,
        },
        legend to name=leg:forecast-legend
    ]
        
        \addplot[ 
            color=hgfgray, 
            densely dashed,
            legend image post style={scale=0.7},
        ] coordinates {
            (0.5, 0.296520752)
            (256, 151.8186248511886)
        };
        \addlegendentry{Ideal scaling};
        
        \addplot[
            color=hgfblue,
            mark=square*,
            mark options={scale=0.7},
            legend image post style={scale=0.7},
        ] table [
            col sep=comma,
            x=mesh_dim_0,
            y=images_s,
            restrict x to domain={0:14}
        ] {data/dp_hero_table_jupiter.csv};
        \addlegendentry{\largejup};

        \addplot[
            color=hgfinformation,
            mark=square*,
            mark options={scale=0.7},
            legend image post style={scale=0.7},
        ] table [
            col sep=comma,
            x=mesh_dim_0,
            y=images_s
        ] {data/dp_hero_table_lumi.csv};
        \addlegendentry{\largelumi};
        
        \addplot[
            color=hgfmatter,
            mark=square*,
            mark options={scale=0.7},
            legend image post style={scale=0.7},
        ] table [
            col sep=comma,
            x=mesh_dim_0,
            y=images_s
        ] {data/dp-scaling-up48-predictive-jupiter.csv};
        
        \addplot[ 
            color=hgfgray, 
            densely dashed,
        ] coordinates {
            (0.5, 0.111664488)
            (256, 57.17221790861947)
        };
        
        \addplot[ 
            color=hgfgray, 
            densely dashed,
            legend image post style={scale=0.7},
        ] coordinates {
            (0.5, 0.564053127)
            (256, 288.7952011624093)
        };

        \node[anchor=south east, align=left, font=\scriptsize] at (axis cs: 6144, 1.0) {$\uparrow$ higher is better};
    \end{axis}
\end{tikzpicture}

%% file: plots/rmse_new.tikz
\begin{tikzpicture}
    \sffamily
    \begin{axis}[
        gbplot,
        width=1\linewidth,
        height=0.7\linewidth,
        ybar,
        bar width=2.4pt,
        xlabel=Variable,
        xtick={1,3,5,7},
        xticklabels={U10, T2M, U850, T850},
        xmin=0.5,
        xmax=7.7,
        ylabel=RMSE,
        ymin=0,
        ymax=1.7,
        error bars/error bar style={
            hgfdarkblue
        }
    ]
        \addplot[
            color=hgfblue, 
            fill=hgfblue80,
            every node near coord/.style={
                font=\tiny,
                color=black
            }
        ] coordinates{
            (1, 0.8434616327285767)
            (3, 0.7236886024475098)
            (5, 1.3109196424484253)
            (7, 0.6442594528198242)
        };

        \addplot[
            color=hgfmatter,
            fill=hgfmatter80,
            every node near coord/.style={
                font=\tiny,
                color=black
            }
        ] coordinates{
            (1, 0.869)
            (3, 0.962)
            (5, 1.303)
            (7, 0.801)
        };

        \addplot[
            color=hgfee,
            fill=hgfee80,
            every node near coord/.style={
                font=\tiny,
                color=black
            }
        ] coordinates{
            (1, 0.786)
            (3, 1.02)
            (5, 1.206)
            (7, 0.773)
        };

        \addplot[
            color=hgfblue,
            fill=hgfblue30,
            postaction={pattern=north east lines, 
            pattern color=hgfblue}, 
            every node near coord/.style={
                font=\tiny,
                color=black
            }
        ] coordinates{
            (1, 1.105748414993286)
            (3, 0.8327817916870117)
            (5, 1.686234712600708)
            (7, 0.8276068568229675)
        };

        \addplot[
            color=hgfmatter,
            fill=hgfmatter30,
            postaction={pattern=north east lines, 
            pattern color=hgfmatter},
            every node near coord/.style={
                font=\tiny,
                color=black
            }
        ] coordinates{
            (1, 0.987)
            (3, 0.993)
            (5, 1.305)
            (7, 0.858)
        };

        \addplot[
            color=hgfee,
            fill=hgfee30,
            postaction={pattern=north east lines, 
            pattern color=hgfee},
            every node near coord/.style={
                font=\tiny,
                color=black
            }
        ] coordinates{
            (1, 0.849)
            (3, 1.011)
            (5, 1.294)
            (7, 0.778)
        };
        
        \node[anchor=north west] at (axis cs:0.4, 1.5) {\tiny $\downarrow$ lower is better};
    \end{axis}
    
\end{tikzpicture}

%% file: plots/crps_new.tikz
\begin{tikzpicture}
    \sffamily
    \begin{axis}[
        gbplot,
        width=1\linewidth,
        height=0.7\linewidth,
        ybar,
        bar width=2.25pt,
        xlabel=Variable,
        xtick={1,3,5,7},
        xticklabels={U10, T2M, U850, T850},
        xmin=0.5,
        xmax=7.7,
        ylabel=CRPS,
        ymin=0,
        ymax=0.9,
        error bars/error bar style={
            hgfdarkblue
        }
    ]
        \addplot[
            color=hgfblue, 
            fill=hgfblue80,
            every node near coord/.style={
                font=\tiny,
                color=black
            }
        ] coordinates{
            (1, 0.4288055896759033)
            (3, 0.3180925548076629)
            (5, 0.6863359212875366)
            (7, 0.3300284445285797)
        };

        \addplot[
            color=hgfmatter,
            fill=hgfmatter80,
            every node near coord/.style={
                font=\tiny,
                color=black
            }
        ] coordinates{
            (1, 0.451)
            (3, 0.454)
            (5, 0.688)
            (7, 0.365)
        };

        \addplot[
            color=hgfee,
            fill=hgfee80,
            every node near coord/.style={
                font=\tiny,
                color=black
            }
        ] coordinates{
            (1, 0.395)
            (3, 0.473)
            (5, 0.637)
            (7, 0.374)
        };

        \addplot[
            color=hgfblue,
            fill=hgfblue30,
            postaction={pattern=north east lines, 
            pattern color=hgfblue}, 
            every node near coord/.style={
                font=\tiny,
                color=black
            }
        ] coordinates{
            (1, 0.5552982091903687)
            (3, 0.3760026097297668)
            (5, 0.8769150376319885)
            (7, 0.4216233193874359)
        };

        \addplot[
            color=hgfmatter,
            fill=hgfmatter30,
            postaction={pattern=north east lines, 
            pattern color=hgfmatter},
            every node near coord/.style={
                font=\tiny,
                color=black
            }
        ] coordinates{
            (1, 0.501)
            (3, 0.467)
            (5, 0.76)
            (7, 0.423)
        };

        \addplot[
            color=hgfee,
            fill=hgfee30,
            postaction={pattern=north east lines, 
            pattern color=hgfee},
            every node near coord/.style={
                font=\tiny,
                color=black
            }
        ] coordinates{
            (1, 0.422)
            (3, 0.457)
            (5, 0.673)
            (7, 0.376)
        };
        
        \node[anchor=north west] at (axis cs:0.4, 0.8) {\tiny $\downarrow$ lower is better};
    \end{axis}
    
\end{tikzpicture}

%% file: plots/spread-skill-ratio_new.tikz
\begin{tikzpicture}
    \sffamily
    \begin{axis}[
        gbplot,
        width=1\linewidth,
        height=0.7\linewidth,
        ybar,
        bar width=2.25pt,
        xlabel=Variable,
        xtick={1,3,5,7},
        xticklabels={U10, T2M, U850, T850},
        xmin=0.4,
        xmax=7.7,
        ylabel=Spread/skill ratio,
        ymin=0,
        ymax=1.9,
        legend style={
            draw=none,
            fill=white,
            legend columns=6,
            font=\scriptsize\bfseries,
            at={(0.5, 1.0)},
            anchor=north,
            nodes={scale=0.8, transform shape}
        },
        legend to name=leg:predictive-performance-legend,
        error bars/error bar style={
            hgfdarkblue
        }
    ]
        \draw[hgfgray, densely dashed] (axis cs:0.0, 1.0) -- (axis cs:8.0, 1.0);
        
        \node[anchor=north west] at (axis cs:3.7, 1.8) {\tiny $\uparrow\downarrow$ over-/under-dispersive};
        
        \addplot[
            color=hgfblue, 
            fill=hgfblue80,
            every node near coord/.style={
                font=\tiny,
                color=black
            }
        ] coordinates{
            (1, 0.7477753162384033)
            (3, 0.8110668659210205)
            (5, 0.7439625859260559)
            (7, 0.761367678642273)
        };
        \addlegendentry{\fancyname-12H}
        
        \addplot[
            color=hgfmatter,
            fill=hgfmatter80,
            every node near coord/.style={
                font=\tiny,
                color=black
            }
        ] coordinates{
            (1,1.043)
            (3, 1.837)
            (5, 1.099)
            (7, 1.243)
        };
        \addlegendentry{IFS ENS-12H}
        
        \addplot[
            color=hgfee,
            fill=hgfee80,
            every node near coord/.style={
                font=\tiny,
                color=black
            }
        ] coordinates{
            (1, 0.93)
            (3, 0.504)
            (5, 0.817)
            (7, 0.623)
        };
        \addlegendentry{FGN-12H}

       \addplot[
            color=hgfblue,
            fill=hgfblue30,
            postaction={pattern=north east lines, 
            pattern color=hgfblue}, 
            every node near coord/.style={
                font=\tiny,
                color=black
            }
        ] coordinates{
            (1, 0.77484530210495)
            (3, 0.8338698744773865)
            (5, 0.7383273243904114)
            (7, 0.8054159283638)
        };
        \addlegendentry{\fancyname-24H}
        
        \addplot[
            color=hgfmatter,
            fill=hgfmatter30,
            postaction={pattern=north east lines, 
            pattern color=hgfmatter},
            every node near coord/.style={
                font=\tiny,
                color=black
            }
        ] coordinates{
            (1,1.01)
            (3, 1.706)
            (5, 1.052)
            (7, 1.178)
        };
        \addlegendentry{IFS ENS-24H}
        
        \addplot[
            color=hgfee,
            fill=hgfee30,
            postaction={pattern=north east lines, 
            pattern color=hgfee},
            every node near coord/.style={
                font=\tiny,
                color=black
            }
        ] coordinates{
            (1, 1.023)
            (3, 0.582)
            (5, 0.928)
            (7, 0.743)
        };
        \addlegendentry{FGN-24H}
    \end{axis}
    
\end{tikzpicture}

%% file: plots/rmse_comparison_heat_wave.tikz
\begin{tikzpicture}[xscale=0.6667]
\node[font=\bfseries\footnotesize] at (3.750,2.700) {European Heat Wave (August 2022)};
\fill[gray!6] (0.000,1.750) rectangle (1.500,2.450);
\node[text=black] at (0.750,2.100) {\scriptsize 3.826};
\fill[gray!6] (1.500,1.750) rectangle (3.000,2.450);
\node[text=black] at (2.250,2.100) {\scriptsize 3.872};
\fill[gray!6] (3.000,1.750) rectangle (4.500,2.450);
\node[text=black] at (3.750,2.100) {\scriptsize 3.977};
\fill[gray!6] (4.500,1.750) rectangle (6.000,2.450);
\node[text=black] at (5.250,2.100) {\scriptsize 4.239};
\fill[gray!6] (6.000,1.750) rectangle (7.500,2.450);
\node[text=black] at (6.750,2.100) {\scriptsize 4.572};
\definecolor{cell1_0}{RGB}{214,232,240}
\fill[cell1_0] (0.000,1.050) rectangle (1.500,1.750);
\node[text=black] at (0.750,1.400) {\scriptsize 3.620};
\definecolor{cell1_1}{RGB}{218,234,242}
\fill[cell1_1] (1.500,1.050) rectangle (3.000,1.750);
\node[text=black] at (2.250,1.400) {\scriptsize 3.674};
\definecolor{cell1_2}{RGB}{228,239,245}
\fill[cell1_2] (3.000,1.050) rectangle (4.500,1.750);
\node[text=black] at (3.750,1.400) {\scriptsize 3.821};
\definecolor{cell1_3}{RGB}{169,208,228}
\fill[cell1_3] (4.500,1.050) rectangle (6.000,1.750);
\node[text=white] at (5.250,1.400) {\scriptsize 3.886};
\definecolor{cell1_4}{RGB}{102,170,206}
\fill[cell1_4] (6.000,1.050) rectangle (7.500,1.750);
\node[text=white] at (6.750,1.400) {\scriptsize 3.969};
\definecolor{cell2_0}{RGB}{212,231,240}
\fill[cell2_0] (0.000,0.350) rectangle (1.500,1.050);
\node[text=black] at (0.750,0.700) {\scriptsize 3.614};
\definecolor{cell2_1}{RGB}{212,231,240}
\fill[cell2_1] (1.500,0.350) rectangle (3.000,1.050);
\node[text=black] at (2.250,0.700) {\scriptsize 3.659};
\definecolor{cell2_2}{RGB}{226,238,244}
\fill[cell2_2] (3.000,0.350) rectangle (4.500,1.050);
\node[text=black] at (3.750,0.700) {\scriptsize 3.810};
\definecolor{cell2_3}{RGB}{162,205,226}
\fill[cell2_3] (4.500,0.350) rectangle (6.000,1.050);
\node[text=white] at (5.250,0.700) {\scriptsize 3.866};
\definecolor{cell2_4}{RGB}{98,168,204}
\fill[cell2_4] (6.000,0.350) rectangle (7.500,1.050);
\node[text=white] at (6.750,0.700) {\scriptsize 3.956};
\definecolor{cell3_0}{RGB}{212,231,240}
\fill[cell3_0] (0.000,-0.350) rectangle (1.500,0.350);
\node[text=black] at (0.750,0.000) {\scriptsize 3.615};
\definecolor{cell3_1}{RGB}{213,231,240}
\fill[cell3_1] (1.500,-0.350) rectangle (3.000,0.350);
\node[text=black] at (2.250,0.000) {\scriptsize 3.661};
\definecolor{cell3_2}{RGB}{224,237,244}
\fill[cell3_2] (3.000,-0.350) rectangle (4.500,0.350);
\node[text=black] at (3.750,0.000) {\scriptsize 3.802};
\definecolor{cell3_3}{RGB}{158,203,225}
\fill[cell3_3] (4.500,-0.350) rectangle (6.000,0.350);
\node[text=white] at (5.250,0.000) {\scriptsize 3.857};
\definecolor{cell3_4}{RGB}{99,168,205}
\fill[cell3_4] (6.000,-0.350) rectangle (7.500,0.350);
\node[text=white] at (6.750,0.000) {\scriptsize 3.959};
\draw[white,line width=0.5pt] (0.000,-0.350) -- (0.000,2.450);
\draw[white,line width=0.5pt] (1.500,-0.350) -- (1.500,2.450);
\draw[white,line width=0.5pt] (3.000,-0.350) -- (3.000,2.450);
\draw[white,line width=0.5pt] (4.500,-0.350) -- (4.500,2.450);
\draw[white,line width=0.5pt] (6.000,-0.350) -- (6.000,2.450);
\draw[white,line width=0.5pt] (7.500,-0.350) -- (7.500,2.450);
\draw[white,line width=0.5pt] (0,-0.350) -- (7.500,-0.350);
\draw[white,line width=0.5pt] (0,0.350) -- (7.500,0.350);
\draw[white,line width=0.5pt] (0,1.050) -- (7.500,1.050);
\draw[white,line width=0.5pt] (0,1.750) -- (7.500,1.750);
\draw[white,line width=0.5pt] (0,2.450) -- (7.500,2.450);
\draw[gray!30] (0,-0.350) rectangle (7.500,2.450);
\node[anchor=east,font=\scriptsize] at (-0.15,2.100) {IFS HRES};
\node[rotate=90,anchor=south,font=\scriptsize] at (-1.45,0.700) {\textbf{BEAST}};
\node[anchor=east,font=\scriptsize] at (-0.15,1.400) {n=32};
\node[anchor=east,font=\scriptsize] at (-0.15,0.700) {n=96};
\node[anchor=east,font=\scriptsize] at (-0.15,0.000) {n=256};
\node[font=\scriptsize] at (0.750,-0.595) {1};
\node[font=\scriptsize] at (2.250,-0.595) {3};
\node[font=\scriptsize] at (3.750,-0.595) {5};
\node[font=\scriptsize] at (5.250,-0.595) {7};
\node[font=\scriptsize] at (6.750,-0.595) {10};
\node[font=\scriptsize] at (3.750,-0.900) {Lead Time [days]};
\definecolor{cb0}{RGB}{79,156,197}
\fill[cb0] (7.950,-0.350) rectangle (8.200,-0.322);
\definecolor{cb1}{RGB}{83,158,198}
\fill[cb1] (7.950,-0.322) rectangle (8.200,-0.294);
\definecolor{cb2}{RGB}{87,161,200}
\fill[cb2] (7.950,-0.294) rectangle (8.200,-0.266);
\definecolor{cb3}{RGB}{91,163,201}
\fill[cb3] (7.950,-0.266) rectangle (8.200,-0.238);
\definecolor{cb4}{RGB}{94,165,203}
\fill[cb4] (7.950,-0.238) rectangle (8.200,-0.210);
\definecolor{cb5}{RGB}{98,168,204}
\fill[cb5] (7.950,-0.210) rectangle (8.200,-0.182);
\definecolor{cb6}{RGB}{102,170,206}
\fill[cb6] (7.950,-0.182) rectangle (8.200,-0.154);
\definecolor{cb7}{RGB}{106,172,207}
\fill[cb7] (7.950,-0.154) rectangle (8.200,-0.126);
\definecolor{cb8}{RGB}{110,175,209}
\fill[cb8] (7.950,-0.126) rectangle (8.200,-0.098);
\definecolor{cb9}{RGB}{114,177,210}
\fill[cb9] (7.950,-0.098) rectangle (8.200,-0.070);
\definecolor{cb10}{RGB}{117,179,211}
\fill[cb10] (7.950,-0.070) rectangle (8.200,-0.042);
\definecolor{cb11}{RGB}{121,182,213}
\fill[cb11] (7.950,-0.042) rectangle (8.200,-0.014);
\definecolor{cb12}{RGB}{125,184,214}
\fill[cb12] (7.950,-0.014) rectangle (8.200,0.014);
\definecolor{cb13}{RGB}{129,186,216}
\fill[cb13] (7.950,0.014) rectangle (8.200,0.042);
\definecolor{cb14}{RGB}{133,189,217}
\fill[cb14] (7.950,0.042) rectangle (8.200,0.070);
\definecolor{cb15}{RGB}{137,191,219}
\fill[cb15] (7.950,0.070) rectangle (8.200,0.098);
\definecolor{cb16}{RGB}{140,193,220}
\fill[cb16] (7.950,0.098) rectangle (8.200,0.126);
\definecolor{cb17}{RGB}{145,196,221}
\fill[cb17] (7.950,0.126) rectangle (8.200,0.154);
\definecolor{cb18}{RGB}{149,198,223}
\fill[cb18] (7.950,0.154) rectangle (8.200,0.182);
\definecolor{cb19}{RGB}{154,201,224}
\fill[cb19] (7.950,0.182) rectangle (8.200,0.210);
\definecolor{cb20}{RGB}{158,203,225}
\fill[cb20] (7.950,0.210) rectangle (8.200,0.238);
\definecolor{cb21}{RGB}{163,205,226}
\fill[cb21] (7.950,0.238) rectangle (8.200,0.266);
\definecolor{cb22}{RGB}{168,208,228}
\fill[cb22] (7.950,0.266) rectangle (8.200,0.294);
\definecolor{cb23}{RGB}{172,210,229}
\fill[cb23] (7.950,0.294) rectangle (8.200,0.322);
\definecolor{cb24}{RGB}{177,213,230}
\fill[cb24] (7.950,0.322) rectangle (8.200,0.350);
\definecolor{cb25}{RGB}{182,215,232}
\fill[cb25] (7.950,0.350) rectangle (8.200,0.378);
\definecolor{cb26}{RGB}{186,217,233}
\fill[cb26] (7.950,0.378) rectangle (8.200,0.406);
\definecolor{cb27}{RGB}{191,220,234}
\fill[cb27] (7.950,0.406) rectangle (8.200,0.434);
\definecolor{cb28}{RGB}{195,222,235}
\fill[cb28] (7.950,0.434) rectangle (8.200,0.462);
\definecolor{cb29}{RGB}{200,225,237}
\fill[cb29] (7.950,0.462) rectangle (8.200,0.490);
\definecolor{cb30}{RGB}{205,227,238}
\fill[cb30] (7.950,0.490) rectangle (8.200,0.518);
\definecolor{cb31}{RGB}{209,229,239}
\fill[cb31] (7.950,0.518) rectangle (8.200,0.546);
\definecolor{cb32}{RGB}{214,232,240}
\fill[cb32] (7.950,0.546) rectangle (8.200,0.574);
\definecolor{cb33}{RGB}{218,234,242}
\fill[cb33] (7.950,0.574) rectangle (8.200,0.602);
\definecolor{cb34}{RGB}{221,236,243}
\fill[cb34] (7.950,0.602) rectangle (8.200,0.630);
\definecolor{cb35}{RGB}{224,237,243}
\fill[cb35] (7.950,0.630) rectangle (8.200,0.658);
\definecolor{cb36}{RGB}{226,238,244}
\fill[cb36] (7.950,0.658) rectangle (8.200,0.686);
\definecolor{cb37}{RGB}{228,239,245}
\fill[cb37] (7.950,0.686) rectangle (8.200,0.714);
\definecolor{cb38}{RGB}{230,241,246}
\fill[cb38] (7.950,0.714) rectangle (8.200,0.742);
\definecolor{cb39}{RGB}{232,242,246}
\fill[cb39] (7.950,0.742) rectangle (8.200,0.770);
\definecolor{cb40}{RGB}{234,243,247}
\fill[cb40] (7.950,0.770) rectangle (8.200,0.798);
\definecolor{cb41}{RGB}{236,244,248}
\fill[cb41] (7.950,0.798) rectangle (8.200,0.826);
\definecolor{cb42}{RGB}{238,245,249}
\fill[cb42] (7.950,0.826) rectangle (8.200,0.854);
\definecolor{cb43}{RGB}{240,247,250}
\fill[cb43] (7.950,0.854) rectangle (8.200,0.882);
\definecolor{cb44}{RGB}{242,248,250}
\fill[cb44] (7.950,0.882) rectangle (8.200,0.910);
\definecolor{cb45}{RGB}{244,249,251}
\fill[cb45] (7.950,0.910) rectangle (8.200,0.938);
\definecolor{cb46}{RGB}{247,250,252}
\fill[cb46] (7.950,0.938) rectangle (8.200,0.966);
\definecolor{cb47}{RGB}{249,251,253}
\fill[cb47] (7.950,0.966) rectangle (8.200,0.994);
\definecolor{cb48}{RGB}{251,253,253}
\fill[cb48] (7.950,0.994) rectangle (8.200,1.022);
\definecolor{cb49}{RGB}{253,254,254}
\fill[cb49] (7.950,1.022) rectangle (8.200,1.050);
\definecolor{cb50}{RGB}{255,255,255}
\fill[cb50] (7.950,1.050) rectangle (8.200,1.078);
\definecolor{cb51}{RGB}{254,253,253}
\fill[cb51] (7.950,1.078) rectangle (8.200,1.106);
\definecolor{cb52}{RGB}{254,251,250}
\fill[cb52] (7.950,1.106) rectangle (8.200,1.134);
\definecolor{cb53}{RGB}{253,250,248}
\fill[cb53] (7.950,1.134) rectangle (8.200,1.162);
\definecolor{cb54}{RGB}{253,248,246}
\fill[cb54] (7.950,1.162) rectangle (8.200,1.190);
\definecolor{cb55}{RGB}{252,246,244}
\fill[cb55] (7.950,1.190) rectangle (8.200,1.218);
\definecolor{cb56}{RGB}{251,244,241}
\fill[cb56] (7.950,1.218) rectangle (8.200,1.246);
\definecolor{cb57}{RGB}{251,242,239}
\fill[cb57] (7.950,1.246) rectangle (8.200,1.274);
\definecolor{cb58}{RGB}{250,241,237}
\fill[cb58] (7.950,1.274) rectangle (8.200,1.302);
\definecolor{cb59}{RGB}{250,239,234}
\fill[cb59] (7.950,1.302) rectangle (8.200,1.330);
\definecolor{cb60}{RGB}{249,237,232}
\fill[cb60] (7.950,1.330) rectangle (8.200,1.358);
\definecolor{cb61}{RGB}{248,235,230}
\fill[cb61] (7.950,1.358) rectangle (8.200,1.386);
\definecolor{cb62}{RGB}{248,233,228}
\fill[cb62] (7.950,1.386) rectangle (8.200,1.414);
\definecolor{cb63}{RGB}{247,232,225}
\fill[cb63] (7.950,1.414) rectangle (8.200,1.442);
\definecolor{cb64}{RGB}{247,230,223}
\fill[cb64] (7.950,1.442) rectangle (8.200,1.470);
\definecolor{cb65}{RGB}{246,228,221}
\fill[cb65] (7.950,1.470) rectangle (8.200,1.498);
\definecolor{cb66}{RGB}{245,226,219}
\fill[cb66] (7.950,1.498) rectangle (8.200,1.526);
\definecolor{cb67}{RGB}{245,224,216}
\fill[cb67] (7.950,1.526) rectangle (8.200,1.554);
\definecolor{cb68}{RGB}{244,222,213}
\fill[cb68] (7.950,1.554) rectangle (8.200,1.582);
\definecolor{cb69}{RGB}{243,219,210}
\fill[cb69] (7.950,1.582) rectangle (8.200,1.610);
\definecolor{cb70}{RGB}{243,217,207}
\fill[cb70] (7.950,1.610) rectangle (8.200,1.638);
\definecolor{cb71}{RGB}{242,214,204}
\fill[cb71] (7.950,1.638) rectangle (8.200,1.666);
\definecolor{cb72}{RGB}{241,212,201}
\fill[cb72] (7.950,1.666) rectangle (8.200,1.694);
\definecolor{cb73}{RGB}{240,209,198}
\fill[cb73] (7.950,1.694) rectangle (8.200,1.722);
\definecolor{cb74}{RGB}{240,207,195}
\fill[cb74] (7.950,1.722) rectangle (8.200,1.750);
\definecolor{cb75}{RGB}{239,204,192}
\fill[cb75] (7.950,1.750) rectangle (8.200,1.778);
\definecolor{cb76}{RGB}{238,201,188}
\fill[cb76] (7.950,1.778) rectangle (8.200,1.806);
\definecolor{cb77}{RGB}{238,199,185}
\fill[cb77] (7.950,1.806) rectangle (8.200,1.834);
\definecolor{cb78}{RGB}{237,196,182}
\fill[cb78] (7.950,1.834) rectangle (8.200,1.862);
\definecolor{cb79}{RGB}{236,194,179}
\fill[cb79] (7.950,1.862) rectangle (8.200,1.890);
\definecolor{cb80}{RGB}{235,191,176}
\fill[cb80] (7.950,1.890) rectangle (8.200,1.918);
\definecolor{cb81}{RGB}{235,189,173}
\fill[cb81] (7.950,1.918) rectangle (8.200,1.946);
\definecolor{cb82}{RGB}{234,186,170}
\fill[cb82] (7.950,1.946) rectangle (8.200,1.974);
\definecolor{cb83}{RGB}{233,184,167}
\fill[cb83] (7.950,1.974) rectangle (8.200,2.002);
\definecolor{cb84}{RGB}{232,181,164}
\fill[cb84] (7.950,2.002) rectangle (8.200,2.030);
\definecolor{cb85}{RGB}{231,178,161}
\fill[cb85] (7.950,2.030) rectangle (8.200,2.058);
\definecolor{cb86}{RGB}{230,175,157}
\fill[cb86] (7.950,2.058) rectangle (8.200,2.086);
\definecolor{cb87}{RGB}{229,171,154}
\fill[cb87] (7.950,2.086) rectangle (8.200,2.114);
\definecolor{cb88}{RGB}{229,168,151}
\fill[cb88] (7.950,2.114) rectangle (8.200,2.142);
\definecolor{cb89}{RGB}{228,165,148}
\fill[cb89] (7.950,2.142) rectangle (8.200,2.170);
\definecolor{cb90}{RGB}{227,162,144}
\fill[cb90] (7.950,2.170) rectangle (8.200,2.198);
\definecolor{cb91}{RGB}{226,159,141}
\fill[cb91] (7.950,2.198) rectangle (8.200,2.226);
\definecolor{cb92}{RGB}{225,155,138}
\fill[cb92] (7.950,2.226) rectangle (8.200,2.254);
\definecolor{cb93}{RGB}{224,152,135}
\fill[cb93] (7.950,2.254) rectangle (8.200,2.282);
\definecolor{cb94}{RGB}{223,149,131}
\fill[cb94] (7.950,2.282) rectangle (8.200,2.310);
\definecolor{cb95}{RGB}{222,146,128}
\fill[cb95] (7.950,2.310) rectangle (8.200,2.338);
\definecolor{cb96}{RGB}{221,143,125}
\fill[cb96] (7.950,2.338) rectangle (8.200,2.366);
\definecolor{cb97}{RGB}{220,140,122}
\fill[cb97] (7.950,2.366) rectangle (8.200,2.394);
\definecolor{cb98}{RGB}{219,136,118}
\fill[cb98] (7.950,2.394) rectangle (8.200,2.422);
\definecolor{cb99}{RGB}{218,133,115}
\fill[cb99] (7.950,2.422) rectangle (8.200,2.450);

\draw[gray!40] (7.950,-0.350) rectangle (8.200,2.450);
\draw (8.200,-0.350) -- (8.280,-0.350);
\node[anchor=west,font=\scriptsize] at (8.320,-0.350) {-15};
\draw (8.200,1.050) -- (8.280,1.050);
\node[anchor=west,font=\scriptsize] at (8.320,1.050) {0};
\draw (8.200,2.450) -- (8.280,2.450);
\node[anchor=west,font=\scriptsize] at (8.320,2.450) {15};
\node[rotate=90,font=\scriptsize] at (9.00,1.050) {\% Diff vs IFS HRES};
\end{tikzpicture}

%% file: plots/rmse_comparison_freeze.tikz
\begin{tikzpicture}[xscale=0.6667]
\node[font=\bfseries\footnotesize] at (3.750,2.700) {European Freeze (December 2022)};
\fill[gray!6] (0.000,1.750) rectangle (1.500,2.450);
\node[text=black] at (0.750,2.100) {\scriptsize 5.730};
\fill[gray!6] (1.500,1.750) rectangle (3.000,2.450);
\node[text=black] at (2.250,2.100) {\scriptsize 5.855};
\fill[gray!6] (3.000,1.750) rectangle (4.500,2.450);
\node[text=black] at (3.750,2.100) {\scriptsize 6.037};
\fill[gray!6] (4.500,1.750) rectangle (6.000,2.450);
\node[text=black] at (5.250,2.100) {\scriptsize 6.196};
\fill[gray!6] (6.000,1.750) rectangle (7.500,2.450);
\node[text=black] at (6.750,2.100) {\scriptsize 6.921};
\definecolor{cell1_0}{RGB}{251,244,241}
\fill[cell1_0] (0.000,1.050) rectangle (1.500,1.750);
\node[text=black] at (0.750,1.400) {\scriptsize 5.875};
\definecolor{cell1_1}{RGB}{249,238,234}
\fill[cell1_1] (1.500,1.050) rectangle (3.000,1.750);
\node[text=black] at (2.250,1.400) {\scriptsize 6.070};
\definecolor{cell1_2}{RGB}{255,254,254}
\fill[cell1_2] (3.000,1.050) rectangle (4.500,1.750);
\node[text=black] at (3.750,1.400) {\scriptsize 6.045};
\definecolor{cell1_3}{RGB}{236,244,248}
\fill[cell1_3] (4.500,1.050) rectangle (6.000,1.750);
\node[text=black] at (5.250,1.400) {\scriptsize 5.969};
\definecolor{cell1_4}{RGB}{137,191,219}
\fill[cell1_4] (6.000,1.050) rectangle (7.500,1.750);
\node[text=white] at (6.750,1.400) {\scriptsize 5.955};
\definecolor{cell2_0}{RGB}{251,244,242}
\fill[cell2_0] (0.000,0.350) rectangle (1.500,1.050);
\node[text=black] at (0.750,0.700) {\scriptsize 5.865};
\definecolor{cell2_1}{RGB}{250,240,237}
\fill[cell2_1] (1.500,0.350) rectangle (3.000,1.050);
\node[text=black] at (2.250,0.700) {\scriptsize 6.045};
\definecolor{cell2_2}{RGB}{252,253,254}
\fill[cell2_2] (3.000,0.350) rectangle (4.500,1.050);
\node[text=black] at (3.750,0.700) {\scriptsize 6.003};
\definecolor{cell2_3}{RGB}{237,245,248}
\fill[cell2_3] (4.500,0.350) rectangle (6.000,1.050);
\node[text=black] at (5.250,0.700) {\scriptsize 5.987};
\definecolor{cell2_4}{RGB}{129,186,216}
\fill[cell2_4] (6.000,0.350) rectangle (7.500,1.050);
\node[text=white] at (6.750,0.700) {\scriptsize 5.894};
\definecolor{cell3_0}{RGB}{251,244,241}
\fill[cell3_0] (0.000,-0.350) rectangle (1.500,0.350);
\node[text=black] at (0.750,0.000) {\scriptsize 5.868};
\definecolor{cell3_1}{RGB}{251,242,239}
\fill[cell3_1] (1.500,-0.350) rectangle (3.000,0.350);
\node[text=black] at (2.250,0.000) {\scriptsize 6.021};
\definecolor{cell3_2}{RGB}{251,253,253}
\fill[cell3_2] (3.000,-0.350) rectangle (4.500,0.350);
\node[text=black] at (3.750,0.000) {\scriptsize 5.989};
\definecolor{cell3_3}{RGB}{238,245,249}
\fill[cell3_3] (4.500,-0.350) rectangle (6.000,0.350);
\node[text=black] at (5.250,0.000) {\scriptsize 5.998};
\definecolor{cell3_4}{RGB}{126,185,215}
\fill[cell3_4] (6.000,-0.350) rectangle (7.500,0.350);
\node[text=white] at (6.750,0.000) {\scriptsize 5.874};
\draw[white,line width=0.5pt] (0.000,-0.350) -- (0.000,2.450);
\draw[white,line width=0.5pt] (1.500,-0.350) -- (1.500,2.450);
\draw[white,line width=0.5pt] (3.000,-0.350) -- (3.000,2.450);
\draw[white,line width=0.5pt] (4.500,-0.350) -- (4.500,2.450);
\draw[white,line width=0.5pt] (6.000,-0.350) -- (6.000,2.450);
\draw[white,line width=0.5pt] (7.500,-0.350) -- (7.500,2.450);
\draw[white,line width=0.5pt] (0,-0.350) -- (7.500,-0.350);
\draw[white,line width=0.5pt] (0,0.350) -- (7.500,0.350);
\draw[white,line width=0.5pt] (0,1.050) -- (7.500,1.050);
\draw[white,line width=0.5pt] (0,1.750) -- (7.500,1.750);
\draw[white,line width=0.5pt] (0,2.450) -- (7.500,2.450);
\draw[gray!30] (0,-0.350) rectangle (7.500,2.450);
\node[anchor=east,font=\scriptsize] at (-0.15,2.100) {IFS HRES};
\node[rotate=90,anchor=south,font=\scriptsize] at (-1.45,0.700) {\textbf{BEAST}};
\node[anchor=east,font=\scriptsize] at (-0.15,1.400) {n=32};
\node[anchor=east,font=\scriptsize] at (-0.15,0.700) {n=96};
\node[anchor=east,font=\scriptsize] at (-0.15,0.000) {n=256};
\node[font=\scriptsize] at (0.750,-0.595) {1};
\node[font=\scriptsize] at (2.250,-0.595) {3};
\node[font=\scriptsize] at (3.750,-0.595) {5};
\node[font=\scriptsize] at (5.250,-0.595) {7};
\node[font=\scriptsize] at (6.750,-0.595) {10};
\node[font=\scriptsize] at (3.750,-0.900) {Lead Time [days]};
\definecolor{cb0}{RGB}{79,156,197}
\fill[cb0] (7.950,-0.350) rectangle (8.200,-0.322);
\definecolor{cb1}{RGB}{83,158,198}
\fill[cb1] (7.950,-0.322) rectangle (8.200,-0.294);
\definecolor{cb2}{RGB}{87,161,200}
\fill[cb2] (7.950,-0.294) rectangle (8.200,-0.266);
\definecolor{cb3}{RGB}{91,163,201}
\fill[cb3] (7.950,-0.266) rectangle (8.200,-0.238);
\definecolor{cb4}{RGB}{94,165,203}
\fill[cb4] (7.950,-0.238) rectangle (8.200,-0.210);
\definecolor{cb5}{RGB}{98,168,204}
\fill[cb5] (7.950,-0.210) rectangle (8.200,-0.182);
\definecolor{cb6}{RGB}{102,170,206}
\fill[cb6] (7.950,-0.182) rectangle (8.200,-0.154);
\definecolor{cb7}{RGB}{106,172,207}
\fill[cb7] (7.950,-0.154) rectangle (8.200,-0.126);
\definecolor{cb8}{RGB}{110,175,209}
\fill[cb8] (7.950,-0.126) rectangle (8.200,-0.098);
\definecolor{cb9}{RGB}{114,177,210}
\fill[cb9] (7.950,-0.098) rectangle (8.200,-0.070);
\definecolor{cb10}{RGB}{117,179,211}
\fill[cb10] (7.950,-0.070) rectangle (8.200,-0.042);
\definecolor{cb11}{RGB}{121,182,213}
\fill[cb11] (7.950,-0.042) rectangle (8.200,-0.014);
\definecolor{cb12}{RGB}{125,184,214}
\fill[cb12] (7.950,-0.014) rectangle (8.200,0.014);
\definecolor{cb13}{RGB}{129,186,216}
\fill[cb13] (7.950,0.014) rectangle (8.200,0.042);
\definecolor{cb14}{RGB}{133,189,217}
\fill[cb14] (7.950,0.042) rectangle (8.200,0.070);
\definecolor{cb15}{RGB}{137,191,219}
\fill[cb15] (7.950,0.070) rectangle (8.200,0.098);
\definecolor{cb16}{RGB}{140,193,220}
\fill[cb16] (7.950,0.098) rectangle (8.200,0.126);
\definecolor{cb17}{RGB}{145,196,221}
\fill[cb17] (7.950,0.126) rectangle (8.200,0.154);
\definecolor{cb18}{RGB}{149,198,223}
\fill[cb18] (7.950,0.154) rectangle (8.200,0.182);
\definecolor{cb19}{RGB}{154,201,224}
\fill[cb19] (7.950,0.182) rectangle (8.200,0.210);
\definecolor{cb20}{RGB}{158,203,225}
\fill[cb20] (7.950,0.210) rectangle (8.200,0.238);
\definecolor{cb21}{RGB}{163,205,226}
\fill[cb21] (7.950,0.238) rectangle (8.200,0.266);
\definecolor{cb22}{RGB}{168,208,228}
\fill[cb22] (7.950,0.266) rectangle (8.200,0.294);
\definecolor{cb23}{RGB}{172,210,229}
\fill[cb23] (7.950,0.294) rectangle (8.200,0.322);
\definecolor{cb24}{RGB}{177,213,230}
\fill[cb24] (7.950,0.322) rectangle (8.200,0.350);
\definecolor{cb25}{RGB}{182,215,232}
\fill[cb25] (7.950,0.350) rectangle (8.200,0.378);
\definecolor{cb26}{RGB}{186,217,233}
\fill[cb26] (7.950,0.378) rectangle (8.200,0.406);
\definecolor{cb27}{RGB}{191,220,234}
\fill[cb27] (7.950,0.406) rectangle (8.200,0.434);
\definecolor{cb28}{RGB}{195,222,235}
\fill[cb28] (7.950,0.434) rectangle (8.200,0.462);
\definecolor{cb29}{RGB}{200,225,237}
\fill[cb29] (7.950,0.462) rectangle (8.200,0.490);
\definecolor{cb30}{RGB}{205,227,238}
\fill[cb30] (7.950,0.490) rectangle (8.200,0.518);
\definecolor{cb31}{RGB}{209,229,239}
\fill[cb31] (7.950,0.518) rectangle (8.200,0.546);
\definecolor{cb32}{RGB}{214,232,240}
\fill[cb32] (7.950,0.546) rectangle (8.200,0.574);
\definecolor{cb33}{RGB}{218,234,242}
\fill[cb33] (7.950,0.574) rectangle (8.200,0.602);
\definecolor{cb34}{RGB}{221,236,243}
\fill[cb34] (7.950,0.602) rectangle (8.200,0.630);
\definecolor{cb35}{RGB}{224,237,243}
\fill[cb35] (7.950,0.630) rectangle (8.200,0.658);
\definecolor{cb36}{RGB}{226,238,244}
\fill[cb36] (7.950,0.658) rectangle (8.200,0.686);
\definecolor{cb37}{RGB}{228,239,245}
\fill[cb37] (7.950,0.686) rectangle (8.200,0.714);
\definecolor{cb38}{RGB}{230,241,246}
\fill[cb38] (7.950,0.714) rectangle (8.200,0.742);
\definecolor{cb39}{RGB}{232,242,246}
\fill[cb39] (7.950,0.742) rectangle (8.200,0.770);
\definecolor{cb40}{RGB}{234,243,247}
\fill[cb40] (7.950,0.770) rectangle (8.200,0.798);
\definecolor{cb41}{RGB}{236,244,248}
\fill[cb41] (7.950,0.798) rectangle (8.200,0.826);
\definecolor{cb42}{RGB}{238,245,249}
\fill[cb42] (7.950,0.826) rectangle (8.200,0.854);
\definecolor{cb43}{RGB}{240,247,250}
\fill[cb43] (7.950,0.854) rectangle (8.200,0.882);
\definecolor{cb44}{RGB}{242,248,250}
\fill[cb44] (7.950,0.882) rectangle (8.200,0.910);
\definecolor{cb45}{RGB}{244,249,251}
\fill[cb45] (7.950,0.910) rectangle (8.200,0.938);
\definecolor{cb46}{RGB}{247,250,252}
\fill[cb46] (7.950,0.938) rectangle (8.200,0.966);
\definecolor{cb47}{RGB}{249,251,253}
\fill[cb47] (7.950,0.966) rectangle (8.200,0.994);
\definecolor{cb48}{RGB}{251,253,253}
\fill[cb48] (7.950,0.994) rectangle (8.200,1.022);
\definecolor{cb49}{RGB}{253,254,254}
\fill[cb49] (7.950,1.022) rectangle (8.200,1.050);
\definecolor{cb50}{RGB}{255,255,255}
\fill[cb50] (7.950,1.050) rectangle (8.200,1.078);
\definecolor{cb51}{RGB}{254,253,253}
\fill[cb51] (7.950,1.078) rectangle (8.200,1.106);
\definecolor{cb52}{RGB}{254,251,250}
\fill[cb52] (7.950,1.106) rectangle (8.200,1.134);
\definecolor{cb53}{RGB}{253,250,248}
\fill[cb53] (7.950,1.134) rectangle (8.200,1.162);
\definecolor{cb54}{RGB}{253,248,246}
\fill[cb54] (7.950,1.162) rectangle (8.200,1.190);
\definecolor{cb55}{RGB}{252,246,244}
\fill[cb55] (7.950,1.190) rectangle (8.200,1.218);
\definecolor{cb56}{RGB}{251,244,241}
\fill[cb56] (7.950,1.218) rectangle (8.200,1.246);
\definecolor{cb57}{RGB}{251,242,239}
\fill[cb57] (7.950,1.246) rectangle (8.200,1.274);
\definecolor{cb58}{RGB}{250,241,237}
\fill[cb58] (7.950,1.274) rectangle (8.200,1.302);
\definecolor{cb59}{RGB}{250,239,234}
\fill[cb59] (7.950,1.302) rectangle (8.200,1.330);
\definecolor{cb60}{RGB}{249,237,232}
\fill[cb60] (7.950,1.330) rectangle (8.200,1.358);
\definecolor{cb61}{RGB}{248,235,230}
\fill[cb61] (7.950,1.358) rectangle (8.200,1.386);
\definecolor{cb62}{RGB}{248,233,228}
\fill[cb62] (7.950,1.386) rectangle (8.200,1.414);
\definecolor{cb63}{RGB}{247,232,225}
\fill[cb63] (7.950,1.414) rectangle (8.200,1.442);
\definecolor{cb64}{RGB}{247,230,223}
\fill[cb64] (7.950,1.442) rectangle (8.200,1.470);
\definecolor{cb65}{RGB}{246,228,221}
\fill[cb65] (7.950,1.470) rectangle (8.200,1.498);
\definecolor{cb66}{RGB}{245,226,219}
\fill[cb66] (7.950,1.498) rectangle (8.200,1.526);
\definecolor{cb67}{RGB}{245,224,216}
\fill[cb67] (7.950,1.526) rectangle (8.200,1.554);
\definecolor{cb68}{RGB}{244,222,213}
\fill[cb68] (7.950,1.554) rectangle (8.200,1.582);
\definecolor{cb69}{RGB}{243,219,210}
\fill[cb69] (7.950,1.582) rectangle (8.200,1.610);
\definecolor{cb70}{RGB}{243,217,207}
\fill[cb70] (7.950,1.610) rectangle (8.200,1.638);
\definecolor{cb71}{RGB}{242,214,204}
\fill[cb71] (7.950,1.638) rectangle (8.200,1.666);
\definecolor{cb72}{RGB}{241,212,201}
\fill[cb72] (7.950,1.666) rectangle (8.200,1.694);
\definecolor{cb73}{RGB}{240,209,198}
\fill[cb73] (7.950,1.694) rectangle (8.200,1.722);
\definecolor{cb74}{RGB}{240,207,195}
\fill[cb74] (7.950,1.722) rectangle (8.200,1.750);
\definecolor{cb75}{RGB}{239,204,192}
\fill[cb75] (7.950,1.750) rectangle (8.200,1.778);
\definecolor{cb76}{RGB}{238,201,188}
\fill[cb76] (7.950,1.778) rectangle (8.200,1.806);
\definecolor{cb77}{RGB}{238,199,185}
\fill[cb77] (7.950,1.806) rectangle (8.200,1.834);
\definecolor{cb78}{RGB}{237,196,182}
\fill[cb78] (7.950,1.834) rectangle (8.200,1.862);
\definecolor{cb79}{RGB}{236,194,179}
\fill[cb79] (7.950,1.862) rectangle (8.200,1.890);
\definecolor{cb80}{RGB}{235,191,176}
\fill[cb80] (7.950,1.890) rectangle (8.200,1.918);
\definecolor{cb81}{RGB}{235,189,173}
\fill[cb81] (7.950,1.918) rectangle (8.200,1.946);
\definecolor{cb82}{RGB}{234,186,170}
\fill[cb82] (7.950,1.946) rectangle (8.200,1.974);
\definecolor{cb83}{RGB}{233,184,167}
\fill[cb83] (7.950,1.974) rectangle (8.200,2.002);
\definecolor{cb84}{RGB}{232,181,164}
\fill[cb84] (7.950,2.002) rectangle (8.200,2.030);
\definecolor{cb85}{RGB}{231,178,161}
\fill[cb85] (7.950,2.030) rectangle (8.200,2.058);
\definecolor{cb86}{RGB}{230,175,157}
\fill[cb86] (7.950,2.058) rectangle (8.200,2.086);
\definecolor{cb87}{RGB}{229,171,154}
\fill[cb87] (7.950,2.086) rectangle (8.200,2.114);
\definecolor{cb88}{RGB}{229,168,151}
\fill[cb88] (7.950,2.114) rectangle (8.200,2.142);
\definecolor{cb89}{RGB}{228,165,148}
\fill[cb89] (7.950,2.142) rectangle (8.200,2.170);
\definecolor{cb90}{RGB}{227,162,144}
\fill[cb90] (7.950,2.170) rectangle (8.200,2.198);
\definecolor{cb91}{RGB}{226,159,141}
\fill[cb91] (7.950,2.198) rectangle (8.200,2.226);
\definecolor{cb92}{RGB}{225,155,138}
\fill[cb92] (7.950,2.226) rectangle (8.200,2.254);
\definecolor{cb93}{RGB}{224,152,135}
\fill[cb93] (7.950,2.254) rectangle (8.200,2.282);
\definecolor{cb94}{RGB}{223,149,131}
\fill[cb94] (7.950,2.282) rectangle (8.200,2.310);
\definecolor{cb95}{RGB}{222,146,128}
\fill[cb95] (7.950,2.310) rectangle (8.200,2.338);
\definecolor{cb96}{RGB}{221,143,125}
\fill[cb96] (7.950,2.338) rectangle (8.200,2.366);
\definecolor{cb97}{RGB}{220,140,122}
\fill[cb97] (7.950,2.366) rectangle (8.200,2.394);
\definecolor{cb98}{RGB}{219,136,118}
\fill[cb98] (7.950,2.394) rectangle (8.200,2.422);
\definecolor{cb99}{RGB}{218,133,115}
\fill[cb99] (7.950,2.422) rectangle (8.200,2.450);
\draw[gray!40] (7.950,-0.350) rectangle (8.200,2.450);
\draw (8.200,-0.350) -- (8.280,-0.350);
\node[anchor=west,font=\scriptsize] at (8.320,-0.350) {-20};
\draw (8.200,1.050) -- (8.280,1.050);
\node[anchor=west,font=\scriptsize] at (8.320,1.050) {0};
\draw (8.200,2.450) -- (8.280,2.450);
\node[anchor=west,font=\scriptsize] at (8.320,2.450) {20};
\node[rotate=90,font=\scriptsize] at (9.00,1.050) {\% Diff vs IFS HRES};
\end{tikzpicture}